\documentclass[a4paper,fleqn,usenatbib]{mnras}

\usepackage{mathptmx}

\usepackage[T1]{fontenc}
\usepackage{ae,aecompl}

\usepackage{graphicx}	
\usepackage{amsmath}	
\usepackage{amssymb}	
\usepackage{cite}
\usepackage{caption}
\usepackage{subfig}
\usepackage{mathtools}

\newcommand{\msun}{\,M$_\odot$}
\newcommand{\rsun}{\,R$_\odot$}

\title[Z Cha as Seen with \textit{TESS}]{The Eclipsing Accreting White Dwarf Z Chameleontis as Seen with \textit{TESS}}

\author[J.M.C. Court et al.]{
J.M.C. Court$^{1}$\thanks{E-mail: James.Court@ttu.edu},
S. Scaringi$^{1}$,
S. Rappaport$^{2}$,
Z. Zhan$^{3}$,
C. Littlefield$^{4}$,
\newauthor N. Castro Segura$^{5}$,
C. Knigge$^{5}$,
T. Maccarone$^{1}$,
M. Kennedy$^{6}$,
P. Szkody$^{7}$,
\newauthor P. Garnavich$^{4}$
\\\\
$^{1}$Department of Physics and Astronomy, Texas Tech University, PO Box 41051, Lubbock, TX 79409, USA\\
$^{2}$Department of Physics, Kavli Institute for Astrophysics and Space Research, M.I.T., Cambridge, MA 02139, USA\\
$^{3}$Department of Earth, Atmospheric, and Planetary Sciences, M.I.T., Cambridge, MA 02139, USA\\
$^{4}$Department of Physics, University of Notre Dame, Notre Dame, IN 46556, USA\\
$^{5}$School of Physics and Astronomy, University of Southampton, Southampton SO17 1BJ, UK\\
$^{6}$Jodrell Bank Centre for Astrophysics, School of Physics and Astronomy, The University of Manchester, Manchester M13 9P, UK\\
$^{7}$Department of Astronomy, University of Washington, Seattle, WA 98195-1580, USA\\
}

\date{Accepted XXX. Received YYY; in original form ZZZ}

\pubyear{2019}

\begin{document}
\label{firstpage}
\pagerange{\pageref{firstpage}--\pageref{lastpage}}
\maketitle

\begin{abstract}
We present results from a study of \textit{TESS} observations of the eclipsing dwarf nova system Z Cha, covering both an outburst and a superoutburst.  We discover that Z Cha undergoes hysteretic loops in eclipse depth - out-of-eclipse flux space in both the outburst and the superoutburst.  The direction that these loops are executed in indicates that the disk size increases during an outburst before the mass transfer rate through the disk increases, placing constraints on the physics behind the triggering of outbursts and superoutbursts.  By fitting the signature of the superhump period in a flux-phase diagram, we find the rate at which this period decreases in this system during a superoutburst for the first time.  We find that the superhumps in this source skip evolutionary stage ``A'' seen during most dwarf nova superoutbursts, even though this evolutionary stage has been seen during previous superoutbursts of the same object.  Finally, O-C values of eclipses in our sample are used to calculate new ephemerides for the system, strengthening the case for a third body in Z Cha and placing new constraints on its orbit.
\end{abstract}




\section{Introduction}

\par Accreting White Dwarfs (AWDs) are astrophysical binary systems in which a star transfers matter to a white dwarf companion via Roche-Lobe overflow. If the white dwarf is not highly magnetised, material which flows through the inner Lagrange point (L1) follows a ballistic trajectory until it impacts the outer edge of an accretion disk, resulting in a bright spot. Material then flows through the disk towards the white dwarf until eventually being accreted. In so-called `dwarf-nova' AWDs, changes in the flow rate through the accretion disk can take place in the form of `outbursts' (e.g. \citealp{Warner_Nova,Meyer_Nova}): dramatic increases in luminosity which persist for timescales of a few days and recur on timescales of weeks to years (e.g. \citealp{Cannizzo_Viscous}). The cause of these outbursts is likely a thermal instability in the disk related to the partial ionisation of hydrogen \citep{Osaki_UGem}. Some dwarf-nova AWDs also undergo `superoutbursts' \citep{vanParadijs_Superoutburst}, longer outbursts believed to be triggered during a normal outburst when the radius of the accretion disk reaches a critical value \citep[e.g.][]{Osaki_Tidal-Thermal}, at which point the disk undergoes a tidal instability and becomes eccentric. Superoutbursts are characterised by the presence of `superhumps' in their optical light curves, which are modulations in luminosity with periods close to the orbital period of the system. These superhumps are believed to be caused by geometric effects within an expanded accretion disk \citep{Horne_Superhump}, and in a simplistic model, can be understood as the precession of an elongated accretion disk. Systems which undergo super outbursts are called SU UMa-type systems after the prototype of these systems, SU Ursae Majoris.
\par Z Chameleontis (\citealt{Mumford_ZCha_Discovery}; hereafter Z Cha) is an SU UMa-type AWD consisting of a white dwarf primary and a red dwarf companion, with a well-constrained orbital period of 1.79\,h \citep{Baptista_ThirdBody}. This system shows both outbursts and superoutbursts and, due to its high inclination angle ($\sim 80^\circ$, see Table \ref{tab:properties}), deep eclipses caused by the red dwarf passing in front of the white dwarf and the accretion disk \citep{Mumford_Eclipses}. Due to this property, the parameters of this system are relatively well-constrained; in Table \ref{tab:properties} we list a number of physical properties of the Z Cha system.  This combination of outbursts, superoutbursts, and eclipses make Z Cha an interesting object with which to probe the evolution of AWDs during these events.
\par In this paper, we report on \textit{TESS} observations of Z Cha taken during two different intervals in 2018.  We describe the observations in Section \ref{sec:obs}, and our results relating to the timing characteristics of the system in Sections \ref{sec:orbital} \& \ref{sec:superhump}.  Finally in \ref{sec:ecl} we perform a population study of the eclipses observed by \textit{TESS}, and discuss how they change over the course of both an outburst and a superoutburst.

\begin{table}
\centering
\begin{tabular}{rll}
\hline
\hline
Parameter&Value&Error\\
\hline
Mass Ratio&0.189&0.004\\
Inclination&$80.44^\circ$&$0.11^\circ$\\
WD Mass&0.803\msun&0.014\msun\\
RD Mass&0.152\msun&0.005\msun\\
WD Radius&0.01046\rsun&0.00017\rsun\\
RD Radius&0.182\rsun&0.002\rsun\\
Separation&0.734\rsun&0.005\rsun\\
\hline
\hline
\end{tabular}
\caption{A table of the physical parameters of the Z Cha system.  Values are from \citet{McAllister_Zeclipses}.  See Section \ref{sec:orbital} for discussion regarding the orbital period of the system.}
\label{tab:properties}
\end{table}

\section{Observations}
\label{sec:obs}

\par The \textit{Transiting Exoplanet Survey Satellite} (\textit{TESS}, \citealp{Ricker_TESS}) is a space-based optical telescope launched in 2018.  The telescope consists of four cameras, each with a field of view (FOV) of $24^\circ\times24^\circ$, resulting in a total telescope FOV of $24^\circ\times96^\circ$.  Each camera consists of $4096\times4096$ pixels, resulting in an effective resolution of $21''$, and is effective at wavelengths of $\sim600$--$1000$\,nm.
\par \textit{TESS}'s primary mission is to perform an all-sky survey to search for transiting exoplanets.  This survey is performed by dividing the sky into a number of `sectors', each of which corresponds to the total field of view of all four cameras.  These sectors overlap near the ecliptic poles; as such, many objects have been or will be observed in multiple sectors.  Each sector is observed for approximately 27 days at a cadence of 2 minutes, and a Full Frame Image (FFI) is returned once every 30 minutes.  Due to telemetry constraints, only images of $\sim16000$ pre-selected $\sim10\times10$ pixel `Postage Stamps' are returned at the optimum 2 minute cadence, creating a Target Pixel File (TPF) for each Postage Stamp.   Simple aperture photometry is applied to each of the TPFs to obtain a barycentred Lightcurve File (LCF) of a selected object within that TPF.  FFIs, TPFs and LCFs from \textit{TESS} are available at the Mikulski Archive for Space Telescopes\footnote{\url{https://archive.stsci.edu/tess/}} (MAST).
\par
\begin{figure*}
    \includegraphics[width=2\columnwidth, trim = 25mm 5mm 32mm 5mm]{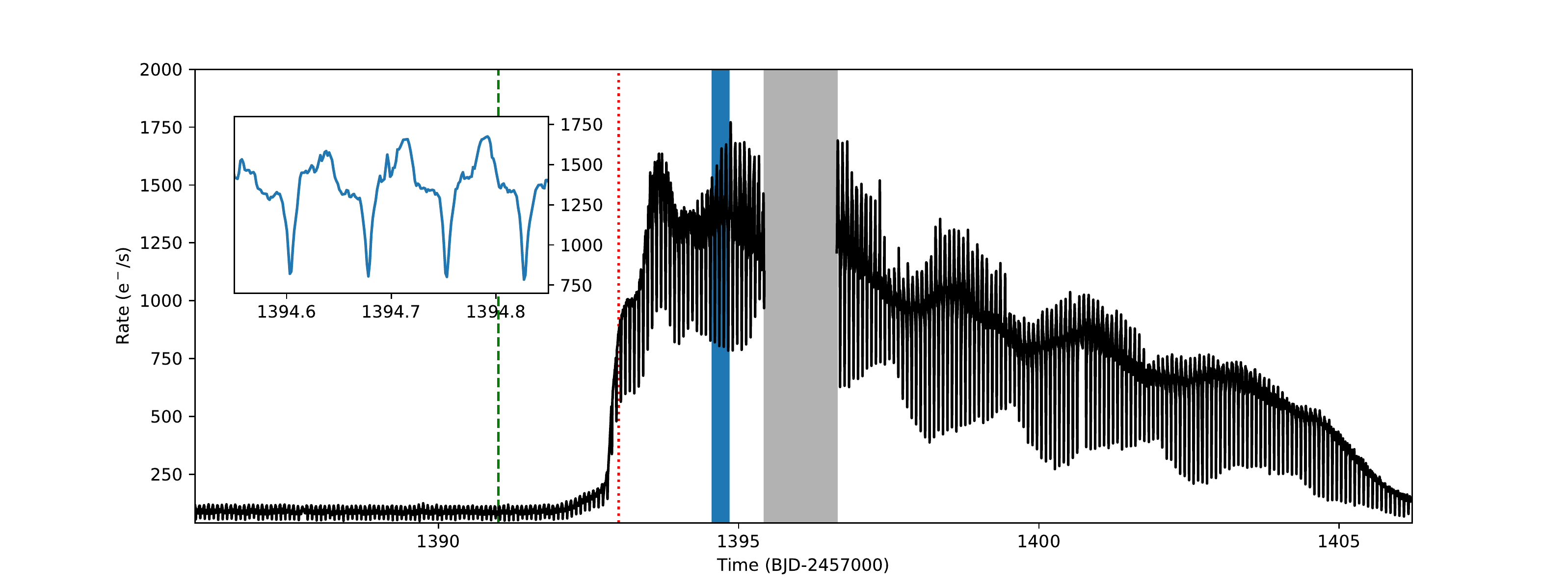}
    \captionsetup{singlelinecheck=off}
    \caption{The full \textit{TESS} lightcurve of Z Cha during Sector 3; inset we plot a zoom in on the region highlighted in blue to show the presence of eclipses.  The data gap centred at about BJD 2458396 (highlighted in grey) is due to the telescope being repointed to downlink data to Earth at this time.  Eclipses at BJDs $\sim2458388$ and $2458401$ occur during smaller data gaps, and hence appear as `missing' eclipses in this lightcurve.  The green dashed line at BJD 2458391 represents a conservative estimate of the start time of the superoutburst which we use to select eclipses which occurred during quiescence (see Section \ref{sec:orbital}).  The red dotted line at BJD 2458393 represents the approximate onset time of superhumps (see Section \ref{sec:superhump})}
   \label{fig:lc3}
\end{figure*}

\begin{figure*}
    \includegraphics[width=2\columnwidth, trim = 25mm 5mm 32mm 5mm]{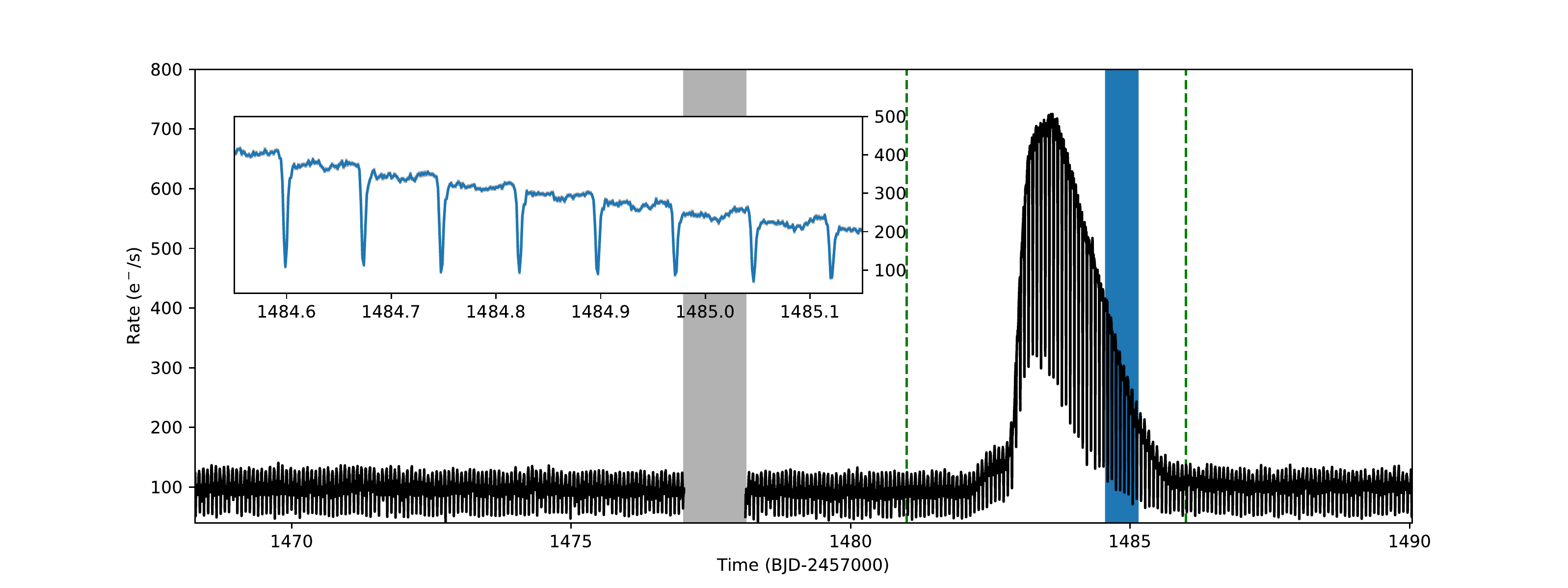}
    \captionsetup{singlelinecheck=off}
    \caption{The full \textit{TESS} lightcurve of Z Cha during Sector 6; inset we plot a zoom in on the region highlighted in blue to show the presence of eclipses.  The data gap centred at about BJD 2458477.5 (highlighted in grey) is due to the telescope being repointed to downlink data to Earth at this time.  Eclipses at BJDs $\sim2458469$ and $2458471$ occur during smaller data gaps, and hence appear as `missing' eclipses in this lightcurve.  The green dashed lines at BJD 2458481 and 2458486 represent conservative estimates of the start and end time of the outburst, which we use to select eclipses which occurred during quiescence (see Section \ref{sec:orbital}).  Note that the vertical scale on this Figure is different to Figure \ref{fig:lc3}.}
   \label{fig:lc6}
\end{figure*}

 Z Cha (Tess Input Catalogue ID 272551828) has been observed during two \textit{TESS} Sectors: Sector 3 between BJDs 2458385--2458406 and Sector 6 between BJDs 2458467--2458490.  We show the LCF-generated Z Cha lightcurves from these sectors in Figures \ref{fig:lc3} and \ref{fig:lc6} respectively.  During Sector 3, Z Cha underwent a superoutburst beginning on BJD 2458391 and persisting until at least the end of Sector 3.  The initial part of this superoutburst took the form of a normal outburst, transitioning to a superoutburst around BJD $\sim2458394$.  During Sector 6, Z Cha underwent a normal outburst beginning on BJD 2458481 and persisting until around BJD 2458486.  Both observations include a significant data gap, in each case caused by the Earth rising above the sun-shade on the spacecraft and contributing significant scattered light\footnote{Data Release Notes (DRNs) on \textit{TESS} Sectors 3 \& 6 can be found at \url{https://archive.stsci.edu/tess/tess_drn.html}}.

\section{Data Analysis}

\label{sec:methods}

\par To analyse \textit{TESS} data from Sectors 3 \& 6, we use our own software libraries\footnote{Available at \url{https://github.com/jmcourt/TTU-libraries}} to extract data from the native \texttt{.fits} LCF files and to create secondary data products such as flux-phase diagrams and power spectra.  As the data in \textit{TESS} LCFs are barycentred, and hence not evenly spaced in time, we do not produce Fourier spectra from the data.  Instead, we analyse timing properties using the Generalised Lomb-Scargle method \citep{Irwin_LombScargle}; a modification on Lomb-Scargle spectral analysis \citep{Lomb_LombScargle,Scargle_LombScargle} which weights datapoints based on their errors.  To extract periods from our Lomb-Scargle spectra, we fit Gaussians in the region of the respective peaks in frequency space.
\par When creating Generalised Lomb-Scargle spectra, we first detrend our data to remove long-term variability and trends such as the evolution of an outburst and a superoutburst.  To perform this detrending, we subtract a value $N_{\rm t}$ from each point in our dataset.  $N_{\rm t}$ is calculated by taking a window of width 0.0744992631\,d, or the orbital period of Z Cha \citep{McAllister_Zeclipses}.  The lower and upper quartiles of the flux values of the data within this window are calculated, and all points outside of the interquartile range are discarded.  $N_{\rm t}$ is then defined as the arithmetic mean of the flux values of the remaining points.  By discarding all values outside of the interquartile range, we remove the flares and eclipses from our detrending process and ensure that our calculated trend is close to describing how the out-of-eclipse rate of the object changes over time.

\subsection{Extracting Eclipses}

\label{sec:extract}

\par
\begin{figure}
    \includegraphics[width=\columnwidth, trim = 2mm 10mm 12mm 5mm]{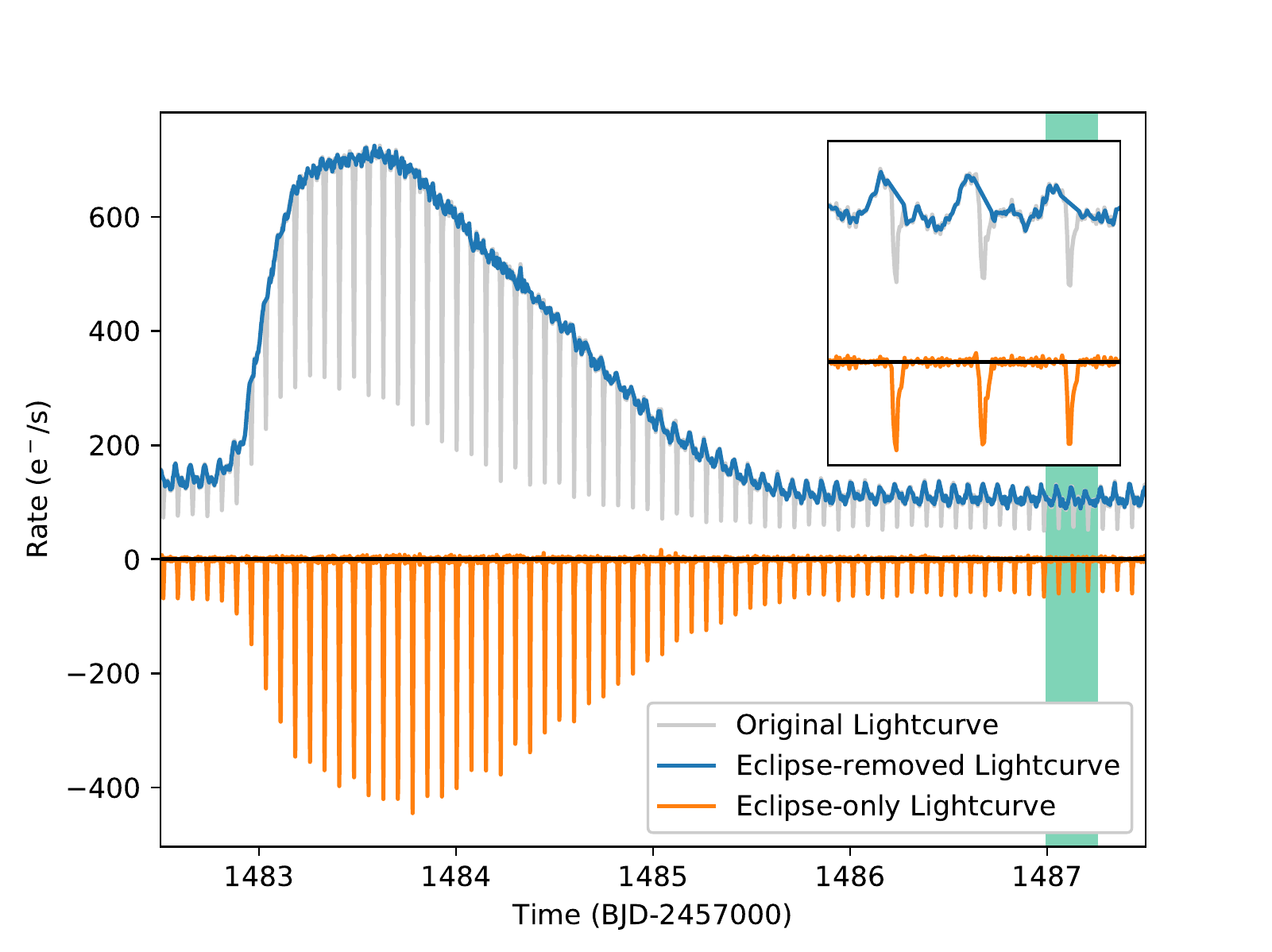}
    \captionsetup{singlelinecheck=off}
    \caption{The results of applying the algorithm we describe in Section \ref{sec:extract} to the portion of the Sector 6 lightcurve of Z Cha containing the outburst.  In light grey we show the original lightcurve, which we have decomposed into an eclipse-removed lightcurve (blue) and an eclipse-only lightcurve (orange).  Inset, we show a zoom to the period during the lightcurves highlighted in green.  Note that, due to the presence of a pre-eclipse brightening caused by the hotspot coming into view (see also Figures \ref{fig:Gfit} \& \ref{fig:folded}), an amount of orbital modulation is still visible in the eclipse-removed lightcurve.}
   \label{fig:isolate_peaks}
\end{figure}

 To study how the eclipse properties varied as a function of time in each sector, we calculated the time of each eclipse minimum assuming an orbital period of 0.0744992631\,d \citep{McAllister_Zeclipses}.  We created a new lightcurve with eclipses removed by removing all data within 0.1 phases ($\sim0.007$\,d) of each eclipse minimum.  We then fit a spline to the remaining data to fill in the gaps using a uniformly-spaced time grid.  This eclipse-free lightcurve could then be subtracted from the original lightcurve to isolate only the eclipse features.  We show some sample resultant lightcurves from this algorithm in Figure \ref{fig:isolate_peaks}.

\par
\begin{figure}
    \includegraphics[width=\columnwidth, trim = 2mm 10mm 12mm 5mm]{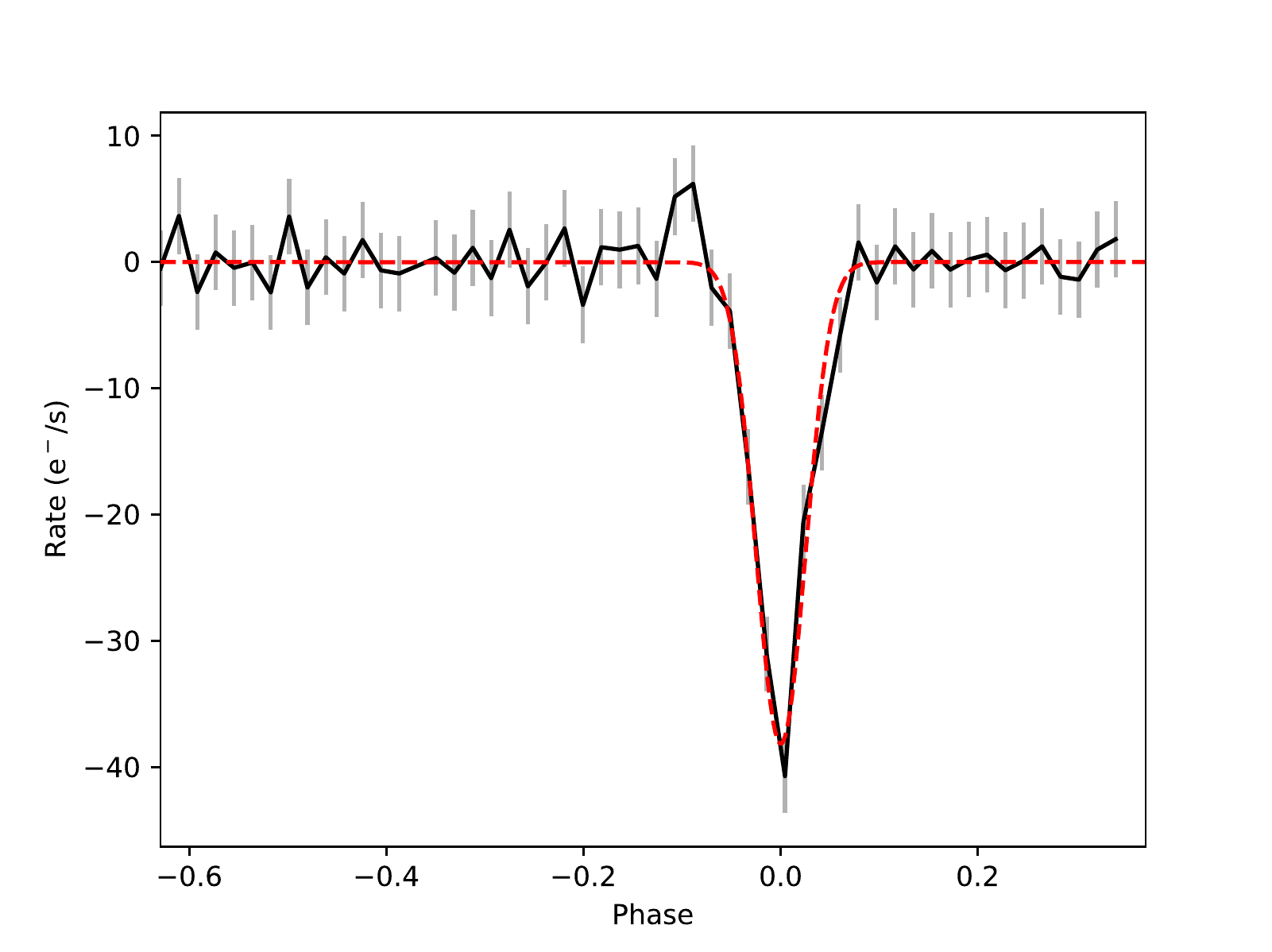}
    \captionsetup{singlelinecheck=off}
    \caption{The lightcurve of a typical eclipse in the eclipse-only lightcurve of Z Cha (black), along with the shape of the Gaussian (red) which we fit to it to extract parameters.}
   \label{fig:Gfit}
\end{figure}

 We split both the eclipse-removed and eclipse-only lightcurve into segments of length equal to one orbital period, such that each segment contains the entirety of a single eclipse.  We fit a Gaussian to each of these segments.  The shape of an eclipse is in general complex, consisting of eclipses of multiple components of the system each with a separate ingress and egress.  Therefore to fit each eclipse, we only fit data less than 0.1 phases before or after the expected eclipse minimum, as the profile of each eclipse is Gaussian-like in this range.  We use this fit to extract a number of parameters for each eclipse, including amplitude $-A$, width $\sigma$ and phase $\phi$; we show an example Gaussian fit in Figure \ref{fig:Gfit}.  We also estimate the out-of-eclipse count rate $\bar{r}$ for each eclipse by taking the median of the corresponding segment of the eclipse-removed lightcurve, resulting in a total of four parameters for each eclipse.

\par A number of small gaps exist in the dataset, generally due to anomalous being manually excluded during the lightcurve processing.  Due to these gaps, some eclipses are partially or completely missing from our sample, and hence a Gaussian fit to these segments of the eclipse-only lightcurve is poorly constrained.  To clean our sample, we remove all eclipses for which the magnitude of any of $A$, $\sigma$ or $\bar{r}$ is less than three times the magnitude of the corresponding error.

\section{Results}

\par In this section we present the results of the analysis we describe in Section \ref{sec:methods}.  First we present our new value for the orbital period of Z Cha, as well as new ephemerides for the system based on fits to historical O-C data.  Then we present our study of the superhump, presenting a new value for its mean frequency and showing that it does not undergo evolutionary stage A \citep{Kato_Shumps1}.  Finally we present our study of the eclipses in this system, showing that hysteresis in max-eclipse-depth - out-of-eclipse-flux space occurs during both the outburst and the superoutburst.

\subsection{Orbital Period}
\label{sec:orbital}

\par
\begin{figure}
    \includegraphics[width=\columnwidth, trim = 2mm 10mm 12mm 5mm]{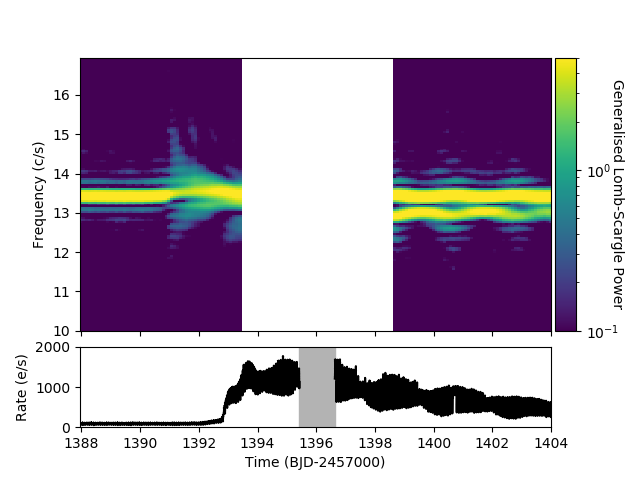}
    \captionsetup{singlelinecheck=off}
    \caption{A dynamic Lomb-Scargle spectrogram of the Sector 3 observation of Z Cha.  Each spectrum corresponds to a 4\,d window of data, which is moved 0.1\,d at a time.  We also show the lightcurve of Sector 3 on the same $x$-axis for comparison.  Note that the dynamic power spectrum is heavily oversampled; using a window size of less than $\sim4$ days does not allow us to resolve the orbital and superhump periods as separate features.}
   \label{fig:LS3}
\end{figure}

\begin{figure}
    \includegraphics[width=\columnwidth, trim = 2mm 10mm 12mm 5mm]{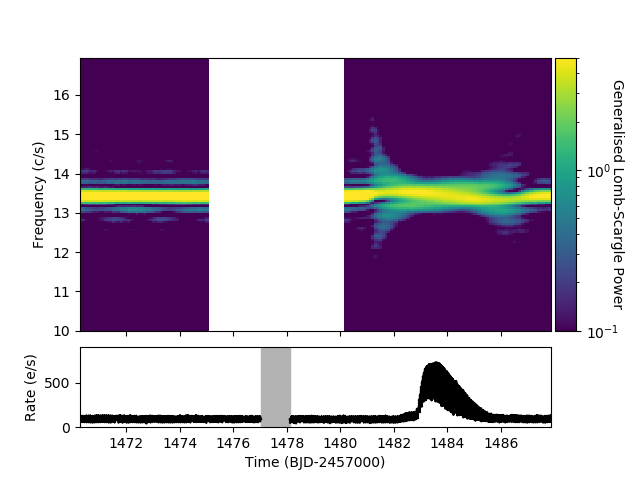}
    \captionsetup{singlelinecheck=off}
    \caption{A dynamic Lomb-Scargle spectrogram of the Sector 6 observation of Z Cha.  Each spectrum corresponds to a 4\,d window of data, which is moved 0.1\,d at a time.  We also show the lightcurve of Sector 6 on the same $x$-axis for comparison.  Note that the dynamic power spectrum is heavily oversampled, consistent with Figure \ref{fig:LS3}}
   \label{fig:LS6}
\end{figure}

 In Figures \ref{fig:LS3} and \ref{fig:LS6}, we show dynamic power spectra constructed from the data of Sectors 3 and 6 respectively, after removing the outburst profile found using the algorithm described in Section \ref{sec:methods}.  In both plots, a strong and constant signal at $\nu_0=13.428(\pm1)$\,c/d can be seen, corresponding to an orbital period of the system $P_{\rm orb}=0.074472(6)$\,d.

\par
\begin{figure}
    \includegraphics[width=\columnwidth, trim = 2mm 12mm 12mm 5mm, clip]{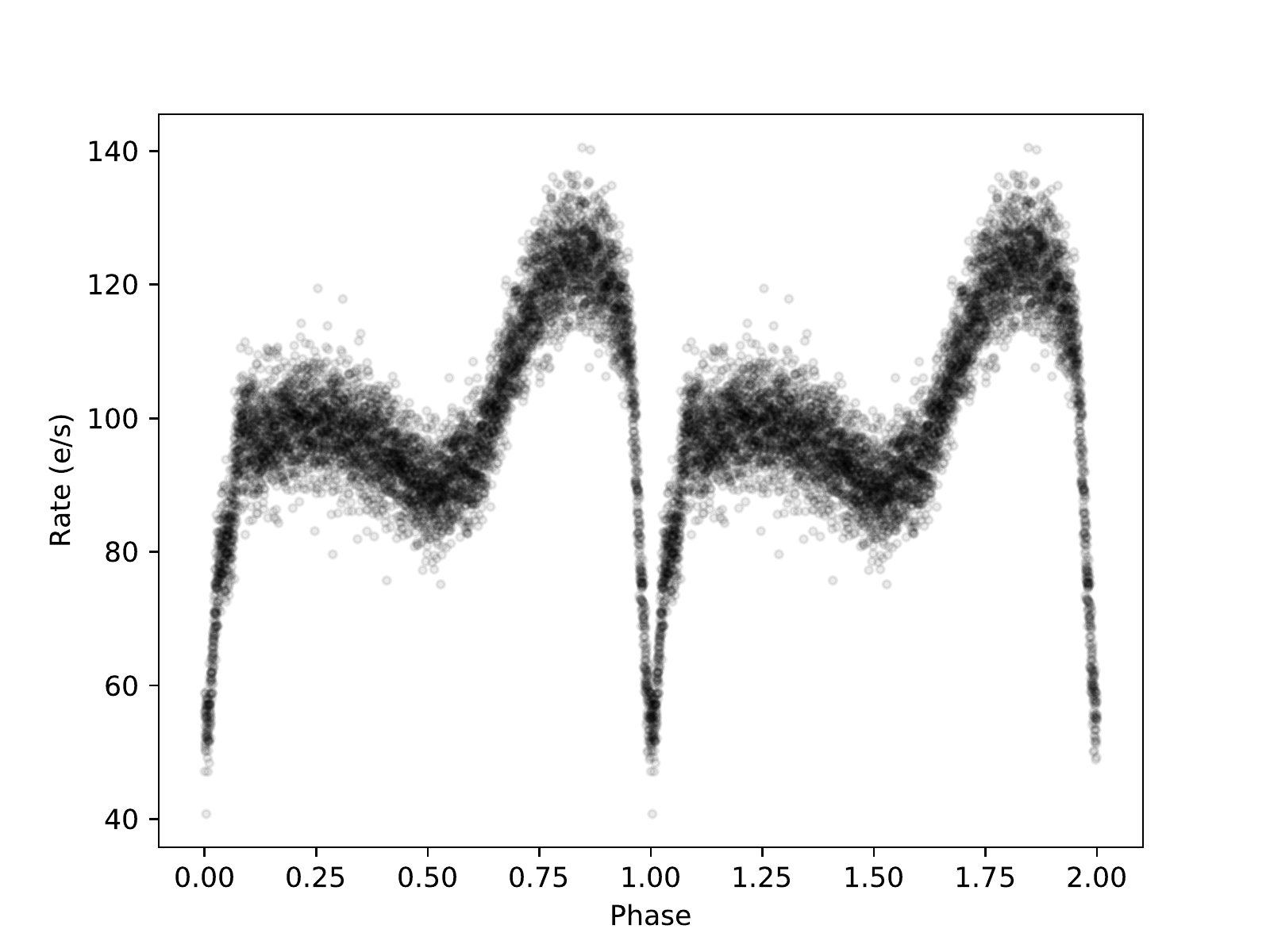}
    \includegraphics[width=\columnwidth, trim = 2mm 00mm 12mm 11mm, clip]{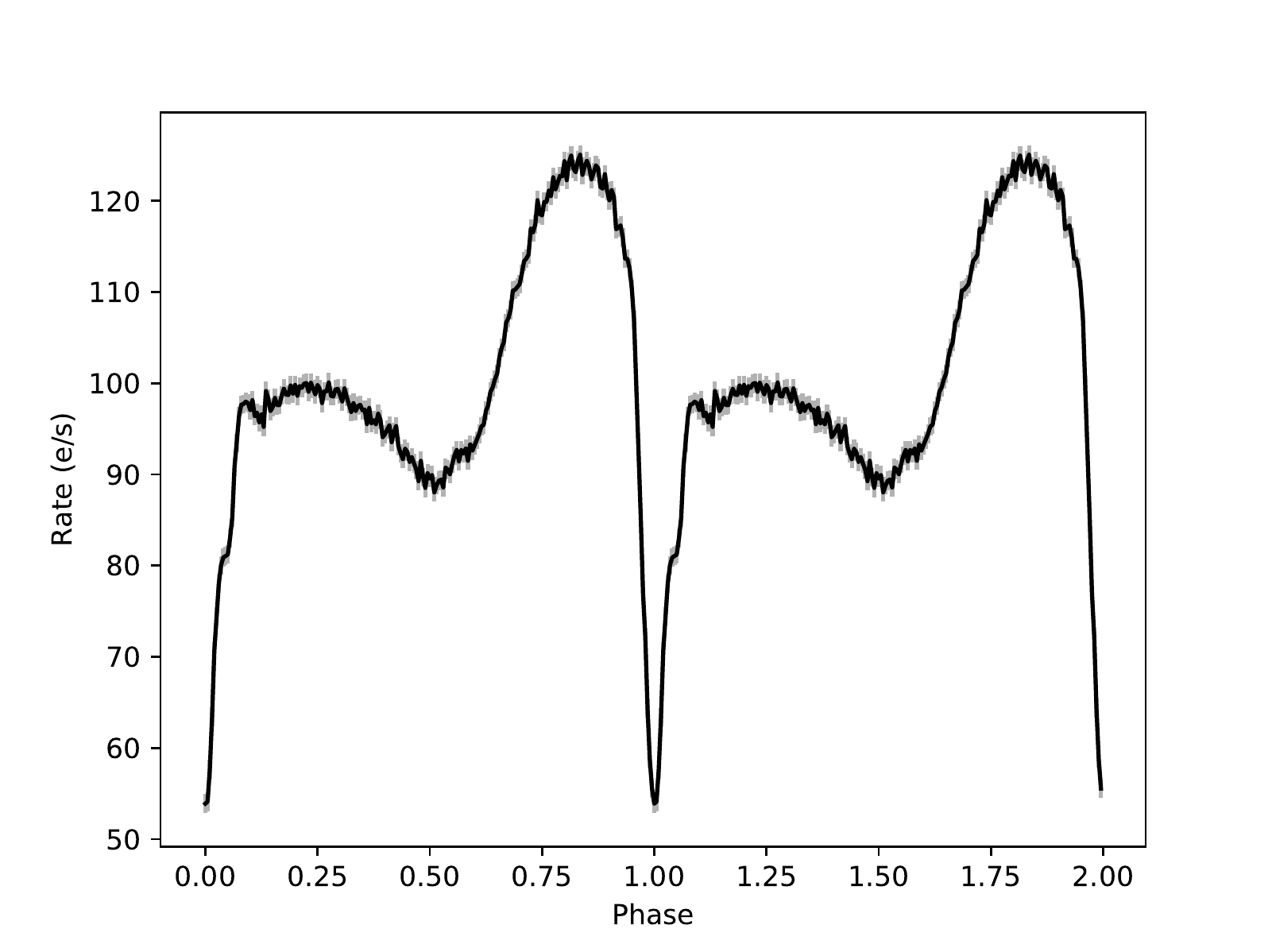}
    \captionsetup{singlelinecheck=off}
    \caption{\textbf{Upper:} The lightcurve of data from Sector 6 before the data gap at BJD $\sim2458477.5$, folded over our measured orbital period of 0.07449953\,d.  We choose this segment of our data to fold here as the out-of-eclipse intensity of Z Cha changes little during this interval, and hence the shape of the average eclipse can be better seen. \textbf{Lower:} the same folded lightcurve, rebinned into phase bins of width 0.005, to better show fine features of the mean cycle such as the egress of the hotspot at $\phi\sim0.1$.}
   \label{fig:folded}
\end{figure}

 We calculate an orbital period $P_{\rm orb}$ for the Z Cha system independently using our dataset so that our value can be used to better constrain the characteristics of the Z Cha system.  Starting with the orbital period of 0.074472\,d indicated by our dynamical power spectrum, we calculated the cycle number (since an arbitrary start time) corresponding to each eclipse minimum during portions of the dataset in which Z Cha was in quiescence.  We then fit a function $t_{\rm min}=P_{\rm orb}N+t_0$ to the eclipse minima times to obtain a value for $P_{\rm orb}$.  We obtain an orbital period of 0.07449953(5), which is slightly longer than the value of $0.0744992631(3)$\,d reported by \citet{McAllister_Zeclipses}.
In Figure \ref{fig:folded}, we show a portion of the lightcurve of Sector 6 folded over our value for the orbital period.

\par We also calculate the O-C values (the difference between observed eclipse time and expected eclipse time according to a given ephemeris) of the eclipses in our sample compared to the linear ephemeris provided by \citealp{Baptista_ThirdBody}, which we reproduce in Table \ref{tab:eph}.  Due to the presence of significant variations of the value of $\phi$ in eclipses during the outburst and superoutburst (discussed in more detail in Section \ref{sec:ecl}), we do not use any eclipses after BJD 2458391 in Sector 3 or between BJDs 2458481 and 2458486 in Sector 6; these times are marked on Figures \ref{fig:lc3} \& \ref{fig:lc6}.  By averaging the O-C values for all eclipses in the remaining periods of quiescence, we find a mean O-C of $-293.3(7)$\,s.

\begin{table}
\centering
\begin{tabular}{l}
\hline
\hline
\citealp{Baptista_ThirdBody} linear ephemeris\\
\hline
$t_{\rm e}=T_0+P_0N$\\
$T_0=2440264.68070(\pm4)$\,d\\
$P_0=0.0744993048(\pm9)$\,d\\
\hline
\hline
\citealp{Robinson_Zeclipses} quadratic ephemeris\\
\hline
$t_{\rm e}=T_0+P_0N+cN^2$\\
$T_0=2440264.63213(\pm9)$\,d\\
$P_0=0.0744992575(\pm24)$\,d\\
$c=3.77(\pm6)$\,d\\
\hline
\hline
\citealp{Baptista_ThirdBody} sinusoid ephemeris\\
\hline
$t_{\rm e}=T_0+P_0N+A\cos(2\pi(N-B)/C)$\\
$T_0=2440264.6817(\pm1)$\,d\\
$P_0=0.074499297(\pm2)$\,d\\
$A=(7.2\pm1.0)\times10^{-4}$\,d\\
$B=(120\pm4)\times10^3$\\
$C=(136\pm7)\times10^3$\\
\hline
\hline
\citealp{Dai_Z_Cha_Dwarf} sinusoid ephemeris\\
\hline
$t_{\rm e}=T_0+P_0N+A\sin(2\pi BN+C)$\\
$T_0=2440265.85(2)$\,d\\
$P_0=0.074499293(4)$\,d\\
$A=(8.8\pm0.1)\times10^{-4}$\,d\\
$B=(2.4\pm3)\times10^{-5}$\\
$C=(0.992\pm7)\times10^3$\\
\hline
\hline
\end{tabular}
\caption{A table of ephemerides for the eclipses of Z Cha for which we simulate O-C values in Figure \ref{fig:OminusC}.  In each case, $t_{\rm e}$ is the expected time of minimum light of the $N^{\mathrm{th}}$ eclipse since some time $T_0$.}
\label{tab:eph}
\end{table}

\par
\begin{figure}
    \includegraphics[width=\columnwidth, trim = 2mm 10mm 12mm 5mm]{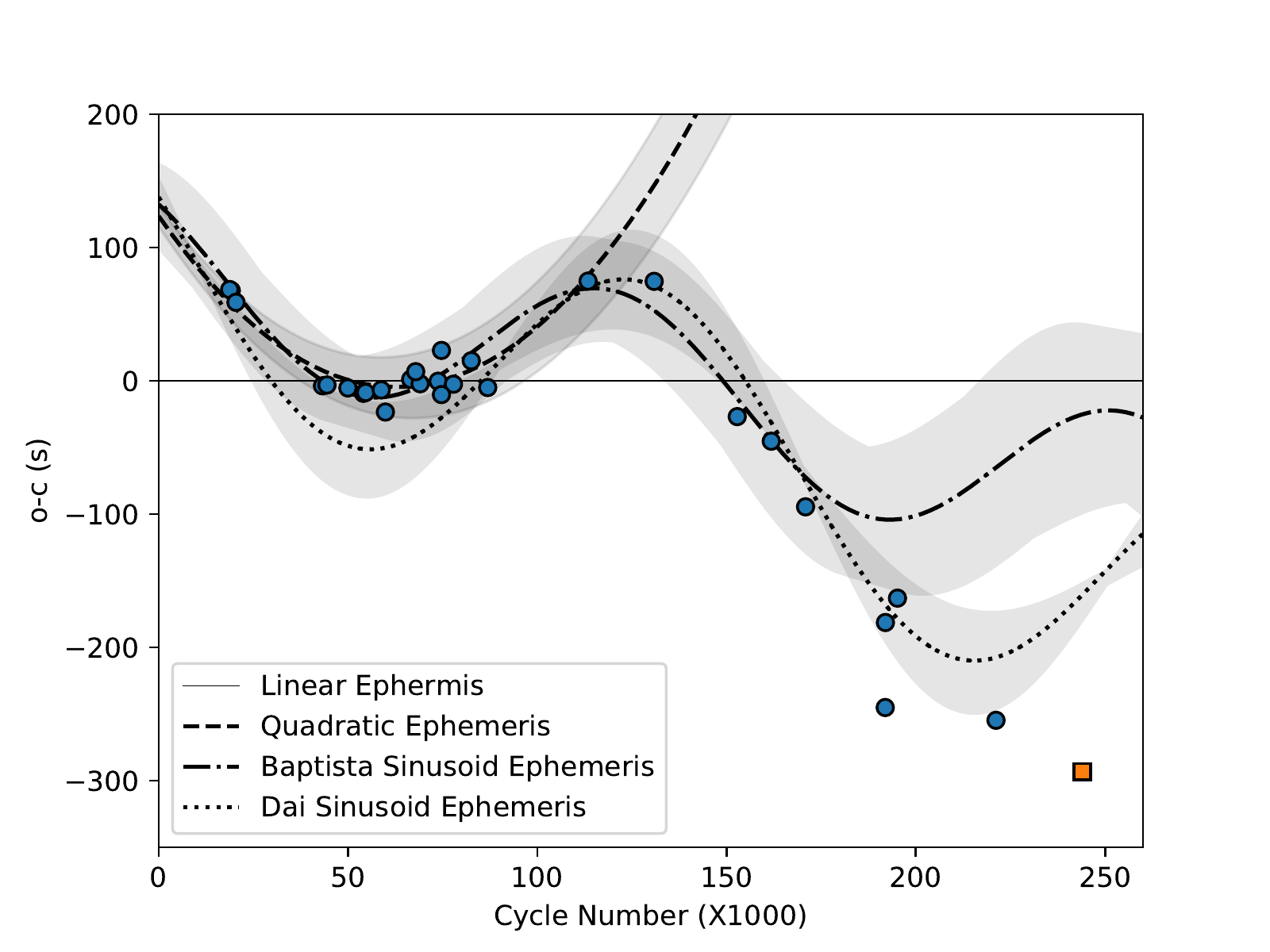}
    \captionsetup{singlelinecheck=off}
    \caption{A plot of the O-C values for reported eclipse times against the linear ephemeris of \citealp{Baptista_ThirdBody} over time.  Blue points are recalculated O-C values for eclipses reported in the literature (see main body of text for sources), whereas the orange square represents the point added by our study.  In black lines we plot the expected O-C values for a number of different ephemerides; see Table \ref{tab:eph} for further details.  The grey area associated with each black line represents the $1\sigma$ confidence interval associated with that ephemeris.}
   \label{fig:OminusC}
\end{figure}

 We show our O-C value in Figure \ref{fig:OminusC} alongside O-C values against the \citealp{Baptista_ThirdBody} ephemeris for eclipse times reported in a number of previous studies \citep{Cook_Zeclipses2,Wood_Zeclipses,Warner_Zeclipses,Honey_Zeclipses,vanAmerongen_Zeclipses,Robinson_Zeclipses,Baptista_ThirdBody,Dai_Z_Cha_Dwarf,Nucita_Zeclipses,Pilarcik_Z_Cha_Period}.  We choose the ephemeris from \citealp{Baptista_ThirdBody} as a reference due to its linearity.

\par A number of previous studies (e.g. \citealp{Baptista_ThirdBody}) have noted an apparently sinusoidal variation in the O-C values of eclipses in Z Cha.  This effect has been seen in other eclipsing AWDs (e.g. \citealp{Bond_Cyclical}), and has variously been attibuted to either a secular process caused by magnetic cycles in the donor star \citep[the Applegate mechanism,][]{Applegate_Mechanism} or the presence of a third body in the system.  In the latter scenario, this third body either periodically transfers angular momentum to the visible components, or causes a periodic shift in the distance, and hence light-travel time, to the visible components.  \citeauthor{Dai_Z_Cha_Dwarf} note that the Applegate mechanism cannot be employed to satisfactorily explain the variations in the orbital period of Z Cha.  Consequently, previous authors (e.g. \citealp{Baptista_ThirdBody,Dai_Z_Cha_Dwarf}) have fit ephemerides with a sinusoidal component to the archival eclipse times of Z Cha, in order to extract the orbital characteristics of the hypothetical third body.  In Figure \ref{fig:OminusC}, we also show the expected O-C values for sinusoidal ephemerides calculated in \citet{Baptista_ThirdBody} and \citet{Dai_Z_Cha_Dwarf}, as well as a quadratic ephemeris presented in \citet{Robinson_Zeclipses}; in each case, our new data point lies far outside the confidence region associated with the respective ephemeris.
\par We calculate a new ephemeris for Z Cha by fitting fitting a function to all the eclipse time data with the form:
\begin{multline}
t_{\rm e}=T_0+P_0N+A\sin\left(\frac{2\pi(t_e-T_1)}{P_1}\right)\\
\approx T_0+P_0N+A\sin\left(\frac{2\pi(T_0+P_0N-T_1)}{P_1}\right)
\label{eq:eph}
\end{multline}
assuming $A\ll P_0N$.
\par We rebin the arrival times into observing runs; periods of time of at most a few days in which the separation between observations is much less than the separation from the next and previous runs.  The error of each of these runs is taken to be the value of its rms scatter; when fitting, we weight each run by the reciprocal square root of this value.  If this error is smaller than 10\,s, we increase it to 10\,s to attempt to account for historical systematic errors.  If a run contains only 1 point, we assume a conservative error of 30\,s.  Even after doing this, the O-C values from historical data must be treated with caution, as a number of different standards for converting times to BJD exist and it is not always clear which is being used by a given author.  These different standards can lead to differences in reported eclipse times on the order of $\sim50$\,s (e.g. \citealp{vanAmerongen_Zeclipses,Eastman_TSt}) and we do not attempt to correct for them here.
\par
\begin{figure}
    \includegraphics[width=\columnwidth, trim = 2mm 10mm 12mm 5mm]{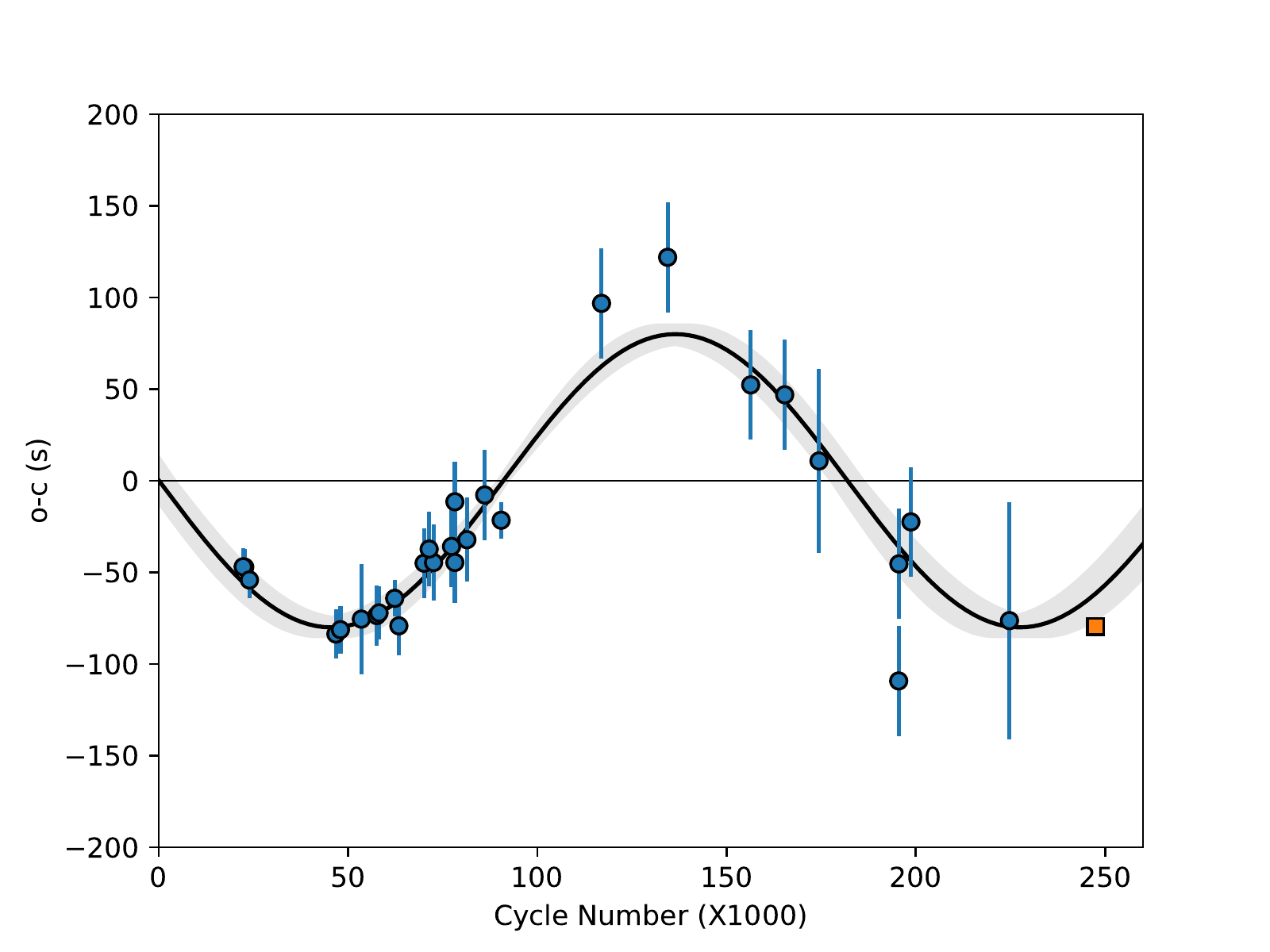}
    \captionsetup{singlelinecheck=off}
    \caption{A plot of the O-C values for reported eclipse times against our new linear ephemeris (see Table \ref{tab:neweph}) over time.  Blue points are recalculated O-C values for eclipses reported in the literature (see main body of text for sources), whereas the orange square represents the point added by our study.  In black lines we plot the expected O-C values from our new sinusoid ephemeris (also in Table \ref{tab:neweph}).  The grey area represents the $1\sigma$ confidence interval associated with our ephemeris.}
   \label{fig:neweph}
\end{figure}

 We present the results of fitting Equation \ref{eq:eph} to our dataset in Table \ref{tab:neweph}, and in Figure \ref{fig:neweph} we show how it fits the O-C taken against a new best-fit linear ephemeris (also given in Table \ref{tab:neweph}).  These values suggest that the orbit of the speculative third body in the Z Cha system has an orbital period of $37.5\pm0.5$\,yr.  We find that the main components of Z Cha likely orbit the centre of mass common to this third body with a semi-major axis of $>82.2\pm5$\,light-seconds (35.4(2) R$_\odot$) depending on the inclination angle.  

\begin{table}
\centering
\begin{tabular}{l}
\hline
\hline
New linear ephemeris\\
\hline
$t_{\rm e}=T_0+P_0N$\\
$T_0=2440000.06088(7)$\,d\\
$P_0=0.0744992878(4)$\,d\\
\hline
\hline
New sinusoid ephemeris\\
\hline
$t_{\rm e}=T_0+P_0N+A\sin\left(\frac{2\pi(T_0+P_0N-T_1)}{P_1}\right)$\\
$T_0=2440000.06088(7)$\,d\\
$P_0=0.0744992878(4)$\,d\\
$A=0.00092(5)$\,d\\
$T_1=2446781\pm140$\,d\\
$P_1=13550\pm438$\,d\\
\hline
\hline
\end{tabular}
\caption{A table of the new best-fit linear and sinusoid ephemerides that we calculate for eclipses in Z Cha.}
\label{tab:neweph}
\end{table}

\subsection{Eclipse Properties}

\label{sec:ecl}

\par As we do not fix the phase when fitting Gaussians to each eclipse, the phase $\phi$ at which minimum light occurs is free to vary slightly from eclipse to eclipse.  We find that, indeed, the minimum-light phase indicated by our fitting does show small but significant variations on orders of 0.01\,$\phi$, especially shortly after the onset of both the outburst and the superoutburst.  This effect has previously been noted by \citealt{Robinson_Zeclipses}, who attribute the phase jump during outbursts to be due to each eclipse consisting of a hybrid event; an eclipse of the hotspot and an eclipse of the disk which cannot be cleanly separated.  The effect has also been noted in a number of other eclipsing AWDs, including V447 Lyrae \citep{Ramsay_PhaseJump} and CRTS J035905.9+175034 \citep{Littlefield_PhaseJump}.


\par In other eclipsing AWDs such as GS Pav \citep{Groot_Pavo} and KIS J192748.53+444724.5 \citep{Scaringi_KIS} it has been shown that, during quiescence, the depth of an eclipse shows a linear correlation with the out of eclipse luminosity.  The gradient of this relationship is very close to unity, which is interpreted as being due to same fraction of the accretion disk flux being obscured during each eclipse.  However, when these systems undergo outbursts, the eclipse depths deviate from this relationship and become shallower.  This is interpreted as being due to a combination of the radial temperature gradient of the disk changing and the physical size of the disk; in each case, the donor star will eclipse a smaller fraction of the integrated luminosity during each eclipse, and hence the eclipses become shallower than would be expected from the $1:1$ relation.
\par
\begin{figure}
    \includegraphics[width=\columnwidth, trim = 2mm 10mm 12mm 5mm]{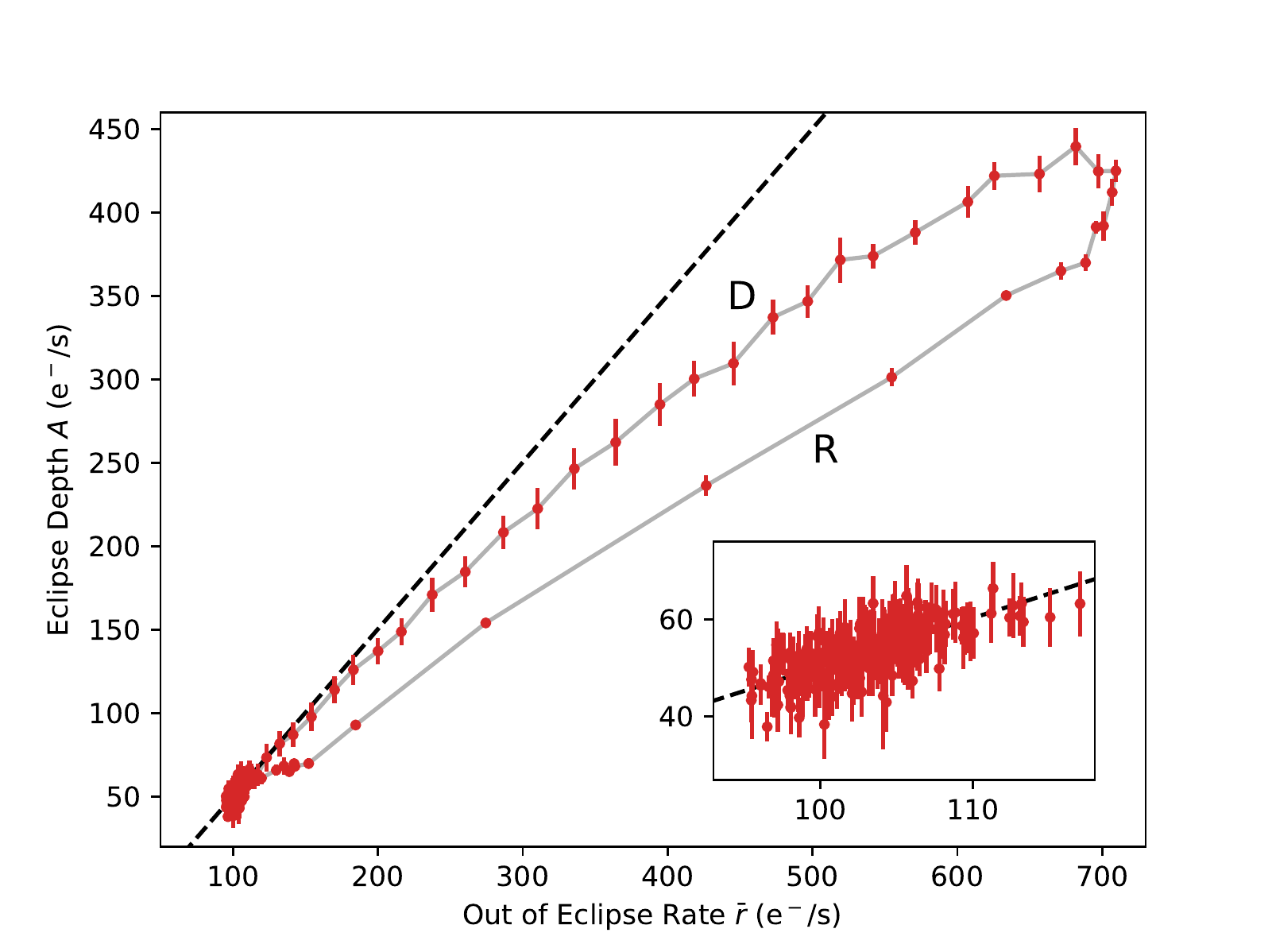}
    \captionsetup{singlelinecheck=off}
    \caption{A plot of eclipse amplitude $A$ against out-of-eclipse rate $\bar{r}$ (an eclipse fraction diagram) for all well-constrained eclipses in the Sector 6 \textit{TESS} observation of Z Cha.  The datapoints are joined sequentially by a grey line to show evidence of hysteresis during the outburst; the track traced by eclipses during the rise (decay) of the outburst is labelled \textbf{R} (\textbf{D}).  The black line is a line with gradient=1 fit to the datapoints corresponding to eclipses during quiescence; inset, we show a zoom to the eclipses used to calculate this line.}
   \label{fig:outburst_ed_ratio6}
\end{figure}

 In Figure \ref{fig:outburst_ed_ratio6} we show a plot of eclipse amplitude $A$ against out-of-eclipse rate $\bar{r}$ (hereafter referred to as an `eclipse fraction diagram') for all eclipses in the Sector 6 \textit{TESS} observation of Z Cha, which includes a normal outburst.  To check that the expected $1:1$ relationship between $A$ and $\bar{r}$ is consistent with our observations, we fit a $y=mx+c$ curve to the portion of this data corresponding to quiescence (e.g. before and after the outburst start and end times, marked with dashed green lines in Figure \ref{fig:lc6}).  We find $m=0.93\pm0.07$.  As this is consistent with 1, we set $m=1$ and fit an $y=x+c$ curve to our data (black line in Figure \ref{fig:outburst_ed_ratio6}), obtaining an $x$-intercept of $49.8(3)$\,e$^-$s$^{-1}$.  This intercept gives the out-of-eclipse flux of the system when the eclipse depth is 0, i.e. there is no disk to eclipse.  As such, this value gives an estimate for the flux of the donor star, and it should be constant across all outbursts of Z Cha.
\par We also show that eclipses during the outburst in Sector 6 exhibit the same behaviour as those seen in other eclipsing AWDs, in that their $A$-values dip below the $1:1$ relationship with $\bar{r}$ during an outburst.  In addition to this, we find strong evidence of hysteresis in this parameter space during the outburst; during the rise of the outburst, eclipses move along a track at lower eclipse depth $A$ (track \textbf{R}, marked in Figure \ref{fig:outburst_ed_ratio6}) and then return to quiescence along a track at higher $A$ (track \textbf{D}).  As the eclipses in our sample are approximately equally spaced in time, the sparsity of points along track \textbf{R} compared to track \textbf{D} indicates that movement along track \textbf{R} is completed in a significantly shorter time.  Hysteresis in this parameter space during a dwarf nova outburst has previously been noted by \citet{Scaringi_KIS}.

\par
\begin{figure}
    \includegraphics[width=\columnwidth, trim = 2mm 10mm 12mm 5mm]{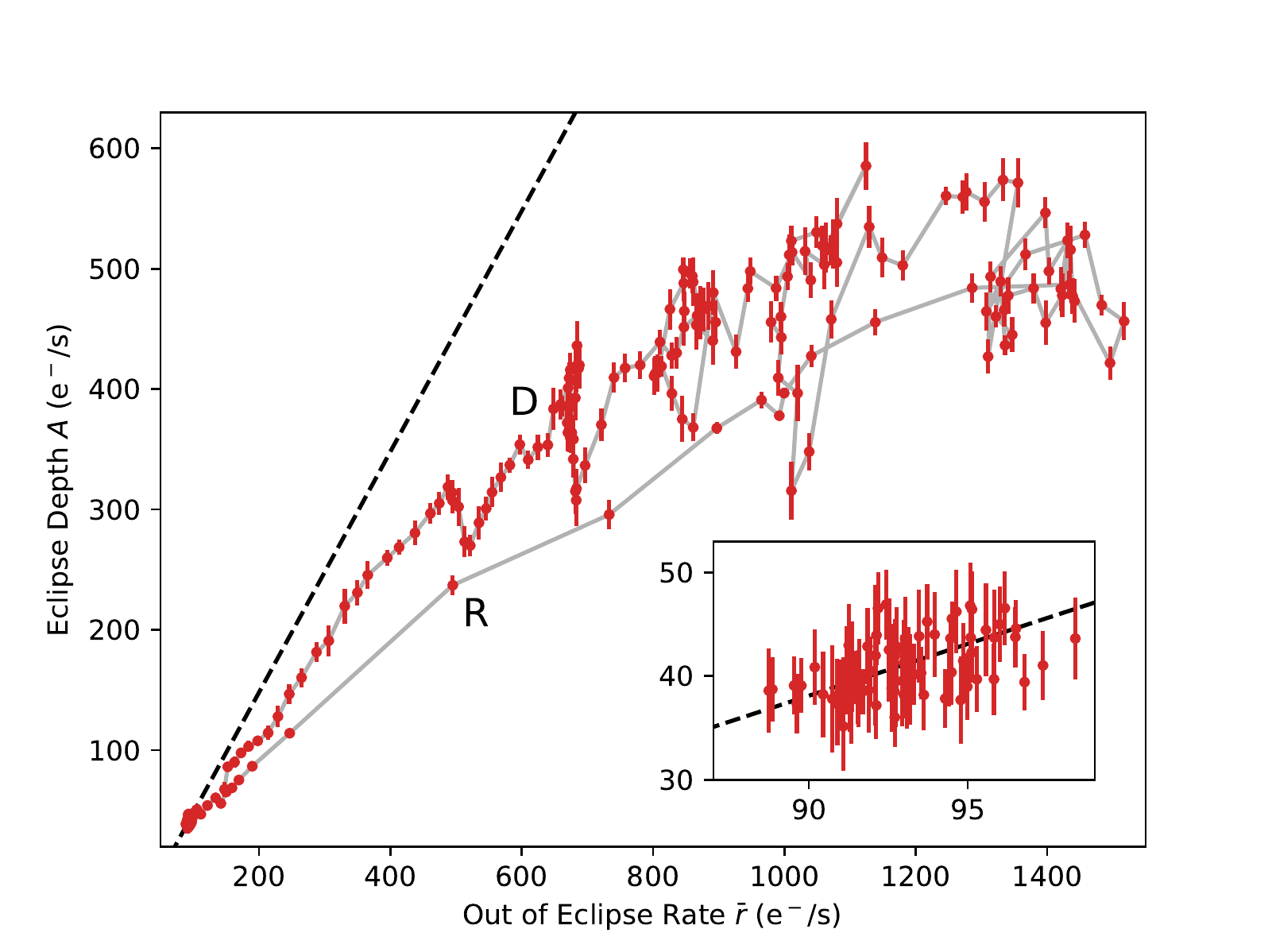}
    \captionsetup{singlelinecheck=off}
    \caption{A plot of eclipse amplitude $A$ against out-of-eclipse rate $\bar{r}$ (an eclipse fraction diagram) for all well-constrained eclipses in the Sector 3 \textit{TESS} observation of Z Cha, which includes the superoutburst.  The datapoints are joined sequentially by a grey line to show evidence of hysteresis during the outburst; the track traced by eclipses during the rise (decay) of the outburst is labelled \textbf{R} (\textbf{D}).  The black line is a line with gradient=1 fit to the datapoints corresponding to eclipses during quiescence; inset, we show a zoom to the eclipses used to calculate this line.}
   \label{fig:outburst_ed_ratio3}
\end{figure}

 In Figure \ref{fig:outburst_ed_ratio3}, we show the eclipse fraction diagram for all eclipses in Sector 3, including the superoutburst.  We again check the $1:1$ correlation between these parameters by fitting a $y=mx+c$ line to the data during quiescence, and obtain a gradient $m=0.7\pm0.2$.  This figure is more than 1\,$\sigma$ lower than 1, although the line is not very well constrained ($c=-28\pm16$\,e$^-$s$^{-1}$) due to the relatively low number of quiescent eclipses in Sector 3 (72) compared to Sector 6 (220).  If we again fix the gradient to 1 and re-fit the line to obtain an $x$-intercept $c=51.9\pm0.3$\,e$^-$s$^{-1}$.  This is significantly different from the $x$-intercept, and hence donor star flux, that we obtain by fitting a straight line to the quiescent periods in Sector 6.  This in turn further suggests that a $1:1$ fit to the data in Sector 3 is not physical.
\par Again we find evidence of hysteresis, confirming that the hysteresis reported by \citet{Scaringi_KIS} during dwarf nova outbursts can also occur during superoutbursts.   Again, the hysteresis generally consists of two tracks: an outbound track \textbf{R} which is completed in relatively little time, and a decay track \textbf{D} which occurs at generally higher values of $A$ than on track \textbf{R}.  The hysteresis in Figure \ref{fig:outburst_ed_ratio3} appears somewhat more complicated than that in Figure \ref{fig:outburst_ed_ratio6}, but some of the excursions to high or low values of $A$ along path \textbf{D} are likely due to the superhump modulation causing us to periodically overestimate $A$.

\subsection{Superhump}
\label{sec:superhump}

\par In Figure \ref{fig:LS3}, a second signal is also visible peaking at a mean frequency of $\sim12.95$\,d$^{-1}$, or a period of $0.07713(\pm7)$\,d.  This signal is only visible during the superoutburst portion of Sector 3, and is absent from all of Sector 6 including the outburst, and so we identify this feature as a `positive superhump'; a modulation of the disk caused by its geometric elongation and subsequent precession after it extends beyond the 3:1 resonance point with the companion star (e.g. \citealp{Wood_Superhumps}).  The superhump begins on BJD $\sim2458393$, less than 3 days before the start of the data gap and at the approximate time of the transition of an outburst into the superoutburst.  Due to the data gap centred around BJD 2458396, and the end of the observation, the early and late-time evolution of the superhump is lost.  Towards the end\footnote{The data for our dynamic Lomb-Scargle periodograms end 2 days before the end of the associated lightcurve data due to our choice of a 4\,d window when producing Lomb-Scargle periodograms.}  of the observation at BJD $\sim2458404$ the amplitude of the superhump begins to weaken significantly, at the same time that the superoutburst is beginning to end (see e.g. Figure \ref{fig:lc3} for comparison).
\par A positive superhump has previously been reported in Z Cha by \citet{Kato_Shumps7} during a 2014 superoutburst of the source.  They calculated the mean period of the superhump as 0.07736(8)\,d during the dominant evolutionary Stage B (see \citealp{Kato_Shumps1}).  Using bootstrapping to estimate errors, we calculate a mean superhump period of $0.07713(\pm7)$\,d from our dataset, somewhat shorter than that calculated by \citeauthor{Kato_Shumps1}.  This discrepancy may in part be due to sampling effects, as the superhump period changes significantly during the course of Stage B.
\par Studies of superhumps in other SU UMa type systems have found `fading' events during superoutburst, in which the system becomes a few tenths of a magnitude fainter for $\sim1$\,d before rebrightening \citep{Littlefield_PhaseJump}.  We see similar features during the superoutburst of Z Cha in the form of a $\sim2$\,d modulation in flux apparent in Figure \ref{fig:lc3}.  However, this modulation occurs very close to the beat period between the orbital period and the mean superhump period ($\sim2.13$\,d), and hence is likely an artefact of this beat.  As the positive superhump is generally interpreted as occurring at the beat frequency of the disk precession and the orbital period \citep{Hellier_Book}, this 2.13\,d period may also be interpreted as the synodic precession period of the elongated accretion disk during superoutburst.
\par
\begin{figure}
    \includegraphics[width=\columnwidth, trim = 2mm 10mm 12mm 5mm]{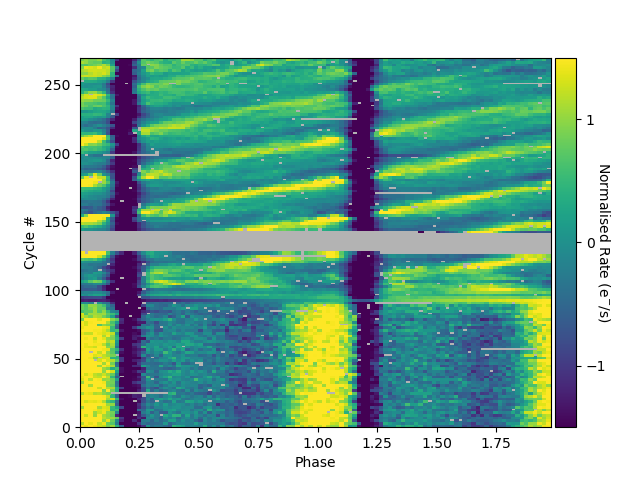}
    \captionsetup{singlelinecheck=off}
    \caption{A flux-phase plot of the Sector 3 lightcurve of Z Cha; a lightcurve which has been cut into segments equal to two times the orbital period and then stacked vertically, to show how lightcurve morphology varies as a function of cycle number.  The brightening associated with the superhump can be seen as diagonal tracks.  The count rates in each orbit have been converted to `normalised rate' by subtracting the mean and dividing by the standard deviation.  Horizontal grey regions correspond to data gaps.}
   \label{fig:flux-phase}
\end{figure}

\begin{figure}
    \includegraphics[width=\columnwidth, trim = 2mm 10mm 12mm 5mm]{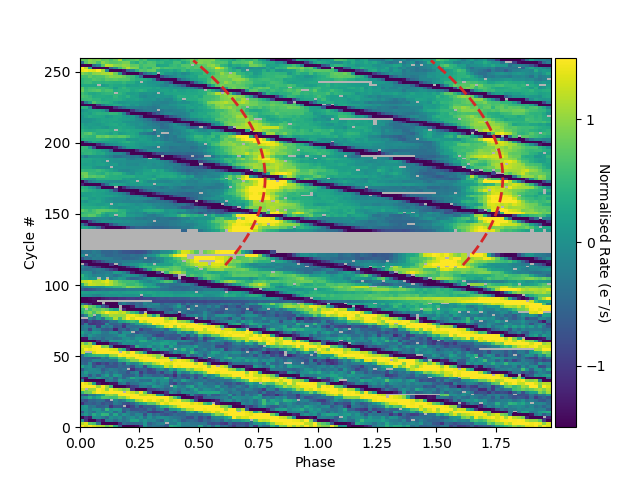}
    \captionsetup{singlelinecheck=off}
    \caption{A flux-phase plot of the Sector 3 lightcurve of Z Cha, folded on the mean superhump period we calculated.  The count rates in each orbit have been converted to `normalised rate' by subtracting the mean and dividing by the standard deviation.  In red we plot the parabola we fit to the central superhump phase as a function of time (see Table \ref{tab:parab}).  Horizontal grey regions correspond to data gaps.}
   \label{fig:flux-phase2}
\end{figure}

 The modulation which causes the superhump can also be readily seen in the lightcurve.  In Figure \ref{fig:flux-phase} we show a flux-phase diagram of the lightcurve from the Sector 3 observation of Z Cha, in which the lightcurve has been folded over the orbital period of the system and then stacked vertically.  This diagram shows how the phase of a periodic or quasiperiodic event changes with time.  The brightening events associated with the superhump can clearly be seen as `diagonal lines' in this plot, arriving slightly later during each orbital period due to the superhump's period being slightly greater than that of the orbit.  In Figure \ref{fig:flux-phase2} we show a similar flux-phase diagram, this time folded over the mean period which we calculated for the superhump.  Including the 10 cycles which occur before the data gap, the superhump appears as a parabola-like shape in this plot; at first it arrives later than its expected time of arrival, but this delay decreases over time and eventually reverses, indicating that the period of the superhump is decreasing over time. 

\section{Discussion}

\subsection{Third Body in the Z Cha System}

\par Sinusoidal O-C modulations have been identified or proposed in a number of additional AWD systems (e.g. \citealp{Bond_Cyclical,Dai_Cyclical,Han_Cyclic}), but the orbital cycles of these systems are generally either poorly sampled or have not been observed in their entirety.  Previous studies have shown that these modulations cannot always be explained by the orbit of a third body (e.g. \citealp{Bianchini_Cyclic}); instead they suggest secular processes such as the Applegate mechanism \citep{Applegate_Mechanism}, in which the donor star is periodically deformed due to its own magnetic activity and this deformation is coupled with a redistribution of angular momentum in the system.  \citet{Dai_Z_Cha_Dwarf} have previously attempted to use the Applegate mechanism to explain the O-C modulation seen in Z Cha.  They calculated the minimum energy required to redistribute sufficient angular momentum in the companion star via this effect to reproduce the observed ampltiude of the oscillation.  They found that the required energy was $\sim1$ order of magnitude greater than the entire energy output of the star over an oscillation period, and hence ruled out the Applegate mechanism as the driver of this oscillation (see also \citealp{Brinkman_Applegate} for a full treatment of the energetics of the Applegate mechanism).  We find a sinusoid amplitude of $79.5\pm4$\,s; as this number is similar to value of $\sim90$\,s obtained by \citeauthor{Dai_Z_Cha_Dwarf}, we find that the energetic output of the donor star is still much too small to explain the O-C modulation via the Applegate mechanism.  As such, our results further strengthen the case for the presence of a low mass third body in Z Cha (see also studies by e.g. \citealp{Baptista_ThirdBody}).
\par If the sinusoidal modulation in O-C is caused by a light-travel time effect, we calculate an orbital period for this body of $37\pm0.5$\,yr, similar to the recent orbital period estimate proposed by \citealt{Dai_Z_Cha_Dwarf} of $\sim32.57$\,yr.    We find the binary mass function of orbit of the tertiary body orbit to be $f(M)=3.2\times10^{-6}$\,\msun.     For a combined mass of the two major components of Z Cha of $0.955$\,\msun \citep{McAllister_Zeclipses}, leads to a brown dwarf mass of $\sim0.015$\,\msun$/\sin i_3$, where $i_3$ is the inclination of the orbital plane of the third body.  Assuming our new estimate for the orbital period, a full orbital cycle of the third body in Z Cha has now been sampled, as can be seen clearly in our fit in Figure \ref{fig:neweph}.  This makes Z Cha one of only very few systems for which this has been achieved (see also e.g. \citealp{Beuermann_CVJupiter} for a similar study on the AWD DP Leonis).
\par However, there are a number of important considerations regarding these results.  The time standard\footnote{In this paper we refer to \textit{TESS}-calculated BJDs, which agree with times calculated using the BJD$_{\rm TDB}$ standard to within 1\,s \citep{Bouma_TESStime}.} employed by prior studies of Z Cha is often unclear during the earlier part of Z Cha's observational history.  The use of these alternative time standards can lead to differences of up to $\sim50$\,s in the reported times of eclipse minima (e.g. \citealp{vanAmerongen_Zeclipses}).  In addition to this, the orbital period we calculate for the third body is similar to the length for which Z Cha has been observed, leading to the possibility that our result may be contaminated by windowing effects.  Future observations of the O-C behaviour of Z Cha will be able to confirm or refute the presence of such a third body in this system.

\subsection{Hysteresis in Eclipse Depth/Out-of-Eclipse Flux Space}

\par It is possible to use the eclipse fraction diagrams, shown in Figures \ref{fig:outburst_ed_ratio6} \& \ref{fig:outburst_ed_ratio3}, to estimate properties of the eclipses in the Z Cha system.  Previous authors have attempted to deconvolve the quiescent eclipses of Z Cha into a series of eclipses of individual components of the system \citep{Wood_Zeclipses,McAllister_Zeclipses}.  This technique leads directly to an estimate of the fraction of the disk flux which is eclipsed during eclipse maximum.  Previous studies \citep[e.g.][]{Wood_Zeclipses,McAllister_Zeclipses} find that the accretion disk is only partially obscured at maximum eclipse depth during quiescence.  However, we find a 1:1 correlation between eclipse depth and out-of-eclipse flux for Z Cha during quiescence in Sector 6.  This suggests that the vast majority of the disk flux is missing during each eclipse at this time, and hence the disk is entirely or almost entirely eclipsed.  To estimate the minimum possible size of the accretion disk, we calculate the circularisation radius of the disk using the formula (e.g. \citealp{Frank_Timescales}):
\begin{equation}
R_{\rm circ}=a(1+q)(0.5-0.227\log{q})^4
\end{equation}
where $a$ is the semi-major axis of the orbit and $q$ is the mass ratio.  Using the values of $q$ and $a$ from \citet{McAllister_Zeclipses} (see Table \ref{tab:properties}), we estimate that the minimum accretion disk radius is $\sim0.170$\,R$_\odot$.  This value is slightly smaller than the red dwarf eclipsor's radius of $0.182$\,R$_\odot$, also taken from \citet{McAllister_Zeclipses}.  As such, the red dwarf in the Z Cha system would be able to totally eclipse the disk if it passes very close to the centre of the disk as seen from Earth.
\par Notably, we were not able to fit a well-constrained 1:1 correlation to eclipse depth and out-of-eclipse flux during the quiescent period before the superoutburst.  This suggests that the maximum eclipse fraction was significantly smaller before the superoutburst than after it in Sector 6.  This in turn suggests that the quiescent disk before the superoutburst had a larger radius than the quiescent disk immediately after the superoutburst.  This finding is consistent with the Tidal-Thermal Instability model of superoutbursts \citep{Osaki_Tidal-Thermal}: in this model, the radius of the accretion disk during quiescence becomes slightly larger after each successive outburst.  Eventually, the quiescent disk radius reaches some critical value, and the next outburst triggers a superoutburst which then resets the quiescent disk radius to some smaller value.  As such, the minimum accretion disk radius should occur during the quiescence immediately after a superoutburst, which is consistent with our finding that the disk radius during \textit{TESS} Sector 6 is likely close to the circularisation radius.
\par The increase in luminosity during AWD outbursts are caused by an increase in both the viscosity (and hence temperature) and the size of the accretion disk.  Using the presence of the hysteresis in the eclipse fraction diagrams which we show in Figures \ref{fig:outburst_ed_ratio6} \& \ref{fig:outburst_ed_ratio3}, it is possible to estimate the sign and magnitude of the response time of the disk to the increase in viscosity or vice versa.  Assuming a static accretion disk (e.g. \citealp{Frank_Timescales}), the temperature $T$ in the disk during outburst as a function of radius $R$ can be expressed as:
\begin{equation}
T(R)=\left(k\frac{\dot{M}}{R^3}\left[1-\sqrt{\frac{R_*}{R}}\right]\right)^{\frac{1}{4}}
\end{equation}
where $\dot{M}$ is the instantaneous accretion rate, $R_*$ is the radius of the compact object and:
\begin{equation}
k=\frac{3GM}{8\pi\sigma}
\end{equation}
which depends only on the mass $M$ of the compact object.  Using the Stefan-Boltzmann equation, the luminosity of one side of such an accretion disk between $R_{\rm in}$ and $R_{\rm out}$ is therefore:
\begin{multline}
L(\dot{M},R_{\rm out})=\frac{1}{2}\int_{R_{\rm in}}^{R_{\rm out}}\sigma 2\pi RT^4(R)\mathrm{\,d}R\\
\propto\dot{M}\left(\frac{2\sqrt{R_*}}{3R_{\rm out}\sqrt{R_{\rm out}}}-\frac{1}{R_{\rm out}}+\frac{1}{3R_{*}}\right)
\end{multline}
assuming that the inner disk radius $R_{\rm in}=R_*$ for an AWD.  The out of eclipse flux from an accretion disk is therefore given by:
\begin{multline}
\Phi_{\rm O}\propto L(\dot{m},r_{\rm disk})\\
\label{eq:rate}
\end{multline}
For a disk with radius $r_{\rm disk}$ and instantaneous mass transfer rate $\dot{m}$.  First of all we assume the simplest possible eclipse, in which the star passes directly in front of the centre of the disk as seen from Earth and the inclination angle of the disk is 90$^\circ$. The maximum eclipse depth can be estimated as the integrated flux over some region $A_0$ of the disk.  $A_0$ can be set equal to the surface of a disk truncated at the radius $r_{\rm ecl}$ of the eclipsing body, unless this is larger than the total flux from the disk:
\begin{equation}
\Phi_{\rm E}\propto
\begin{dcases}
L(\dot{m},r_{\rm disk})&\text{if }r_{\rm ecl}>r_{\rm disk}\\
L(\dot{m},r_{\rm ecl}) &\text{otherwise}
\end{dcases}
\label{eq:depth}
\end{equation}
\par In true astrophysical eclipses the inclination angle of the disk must be small, and the eclipsed region corresponds to a portion $A_{\rm i}$ of the disk highlighted in Figure \ref{fig:highlight_bit}.  In the limit where the disk radius tends to infinity and its inclination tends to 0$^\circ$, the factor by which we underestimate the eclipsed area of the disk tends to:
\begin{figure}
    \includegraphics[width=\columnwidth, trim = 2mm 10mm 12mm 5mm]{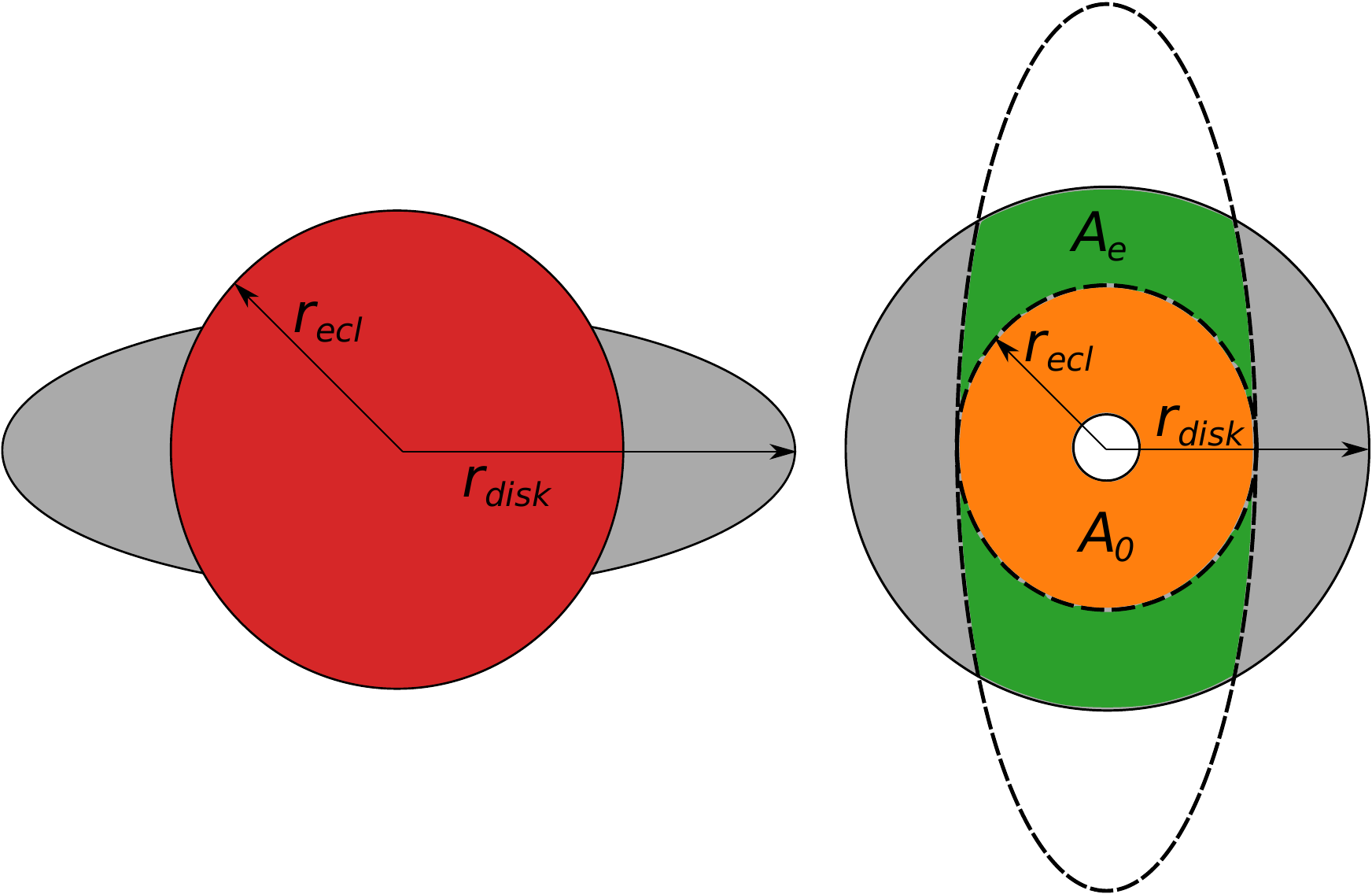}
    \captionsetup{singlelinecheck=off}
    \caption{The fractional projected area of a sphere (red) on the surface of an inclined disk (grey) is equal to the fractional projected area of an ellipse onto a face-on disk (right), where the eccentricity of the ellipse in the latter case is determined by the inclination of the disk in the former.  In our approximation, we assume that the covered region of the disk at peak eclipse ($A_0$ in orange) is equal to a circle with the radius of the eclipser.  The true covered region of the disk is $A_{\rm i}=A_0+A_{\rm e}$, so we underestimate our eclipse depths by a value equal to the integrated flux over $A_{\rm e}$ (in green).}
   \label{fig:highlight_bit}
\end{figure}
\begin{equation}
\frac{A_{\rm i}}{A_0}\xrightarrow{r_{\rm disk}\to\infty,\,i\to0}\frac{4}{\pi}\frac{r_{\rm disk}}{r_{\rm ecl}}
\end{equation}

\par In the Tidal-Thermal Instability model, the radius of the accretion disk during quiescence cannot exceed the radius of a Keplerian 3:1 resonance with the orbital frequency \citep{Osaki_Tidal-Thermal}.  We estimate a value for this radius in Z Cha to be $\sim0.33$\,\rsun, and hence $r_{\rm disk}/r_{\rm ecl}\lesssim2.2$.  As such, our toy model underestimates the eclipsed area of the disk by less than a factor of 3.  As the temperature profile of the disk leads to a surface brightness that decreases with increasing $r$, our model underestimates the eclipsed flux from the disk by even less, and our approximation is valid.
\par
\begin{figure}
    \includegraphics[width=\columnwidth, trim = 2mm 0mm 2mm 5mm]{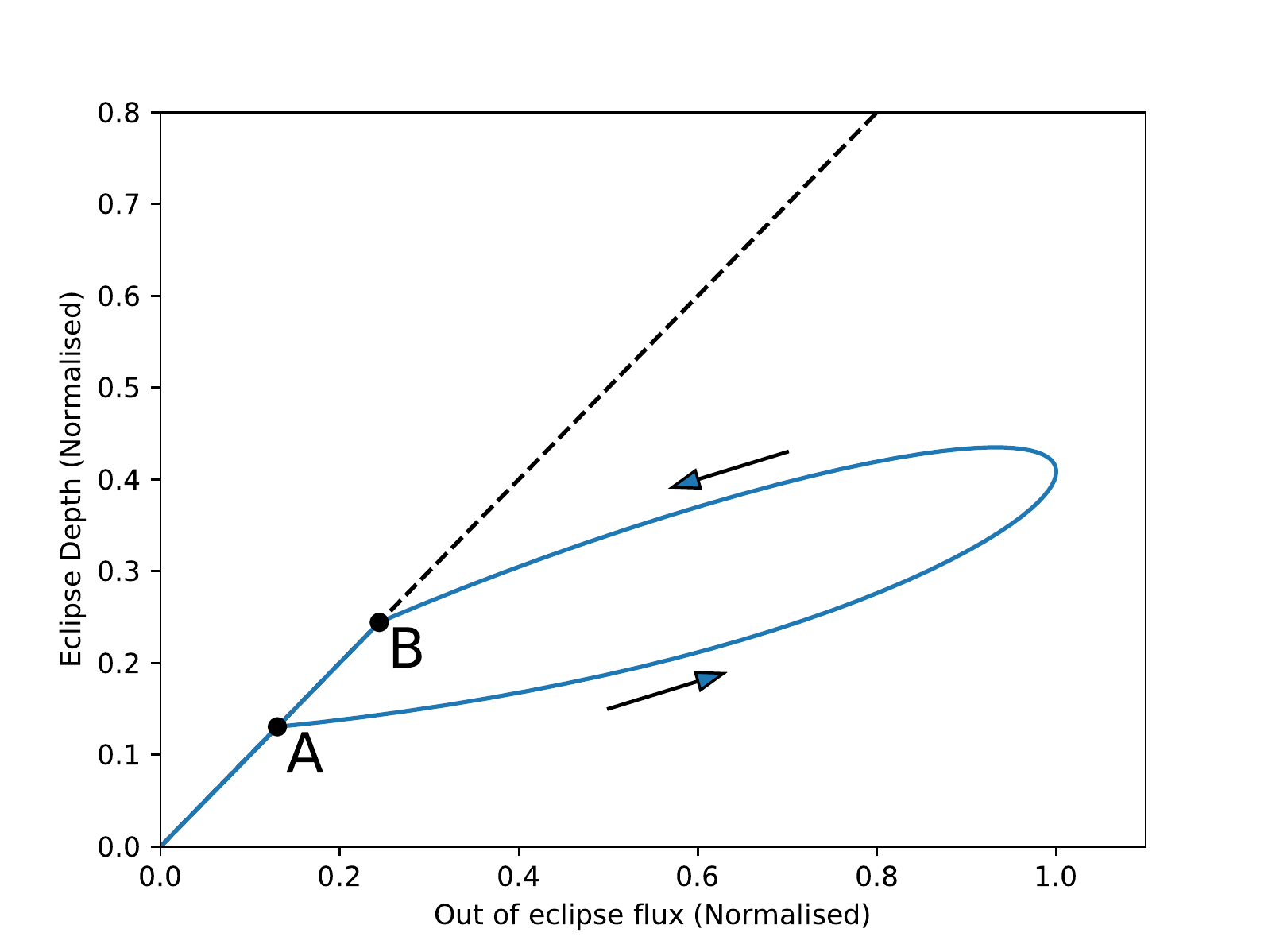}
    \includegraphics[width=\columnwidth, trim = 2mm 0mm 2mm 12mm, clip]{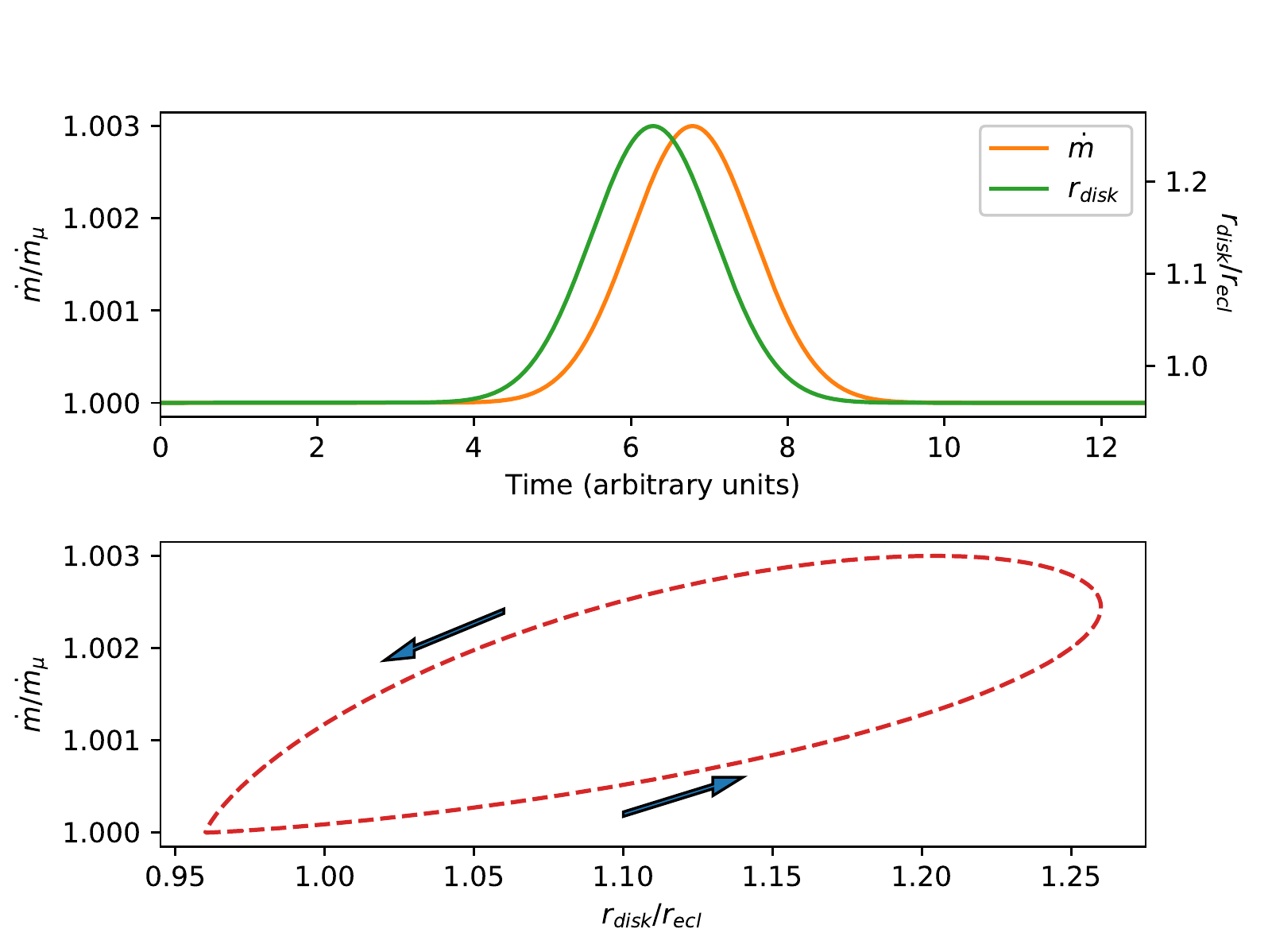}
    \captionsetup{singlelinecheck=off}
    \caption{\textbf{Top Panel:} A plot of eclipse depth against out-of-eclipse flux (an eclipse fraction diagram) for a system modelled using equations \ref{eq:rate} \& \ref{eq:depth}, with Gaussian input signals of the form $\dot{m}=A_1\exp(B_1(t-C)^2)$ and $r_{\rm disk}=A_2\exp(B_2(t-C-\phi)^2)$.  All values have been normalised by the maximum out-of-eclipse flux.  In order for the hysteretic loop in this diagram to be executed in an anticlockwise direction (to mimic the data from Z Cha), $-\pi<\phi<0$ .  \textbf{A} and \textbf{B} mark the points in the diagram at which the object leaves and returns to the 1:1 line in this diagram respectively.  \textbf{Lower Panels:} plots showing the forms of the input signals $\dot{m}(t)$ and $r_{\rm disk}(t)$ used to generate the top panel.  The accretion rate $\dot{m}$ is normalised by the mean accretion rate $\dot{m}_\mu$.}
   \label{fig:toymodel}
\end{figure}

 Using Equations \ref{eq:rate} \& \ref{eq:depth}, it is possible to take two signals $\dot{m}(t)$ and $r_{\rm disk}(t)$ and produce the expected eclipse fraction diagrams, up to some constant.  We take the case in which $\dot{m}(t)$ and $r_{\rm disk}(t)$ have the same functional form, but $r_{\rm disk}$ instead depends on $t-\phi$ for some constant $\phi$.  This way we can model how the eclipse depth diagram would appear for an accretion disk in which an increase in radius lags an increase in accretion rate or vice versa.  We show such an eclipse depth diagram for a Gaussian input signal of the form $\dot{m}=A\exp(B(t-C)^2)$ in Figure \ref{fig:toymodel}.

\par The general behaviour of this modelled eclipse fraction diagram is similar to what we see in the real data from the normal outburst of Z Cha; at low luminosities, eclipse depth and out-of-eclipse flux follow a 1:1 relationship.  At some point (labelled \textbf{A} in Figure \ref{fig:toymodel}), the radius of the disk significantly exceeds the radius of the eclipsing object, and eclipse depth becomes smaller than out-of-eclipse flux.  As the object evolves, it executes a loop in this parameter space below the $x=y$ line, before returning to that line at a point \textbf{B}.
\par For all input signals, we find that the loop in the eclipse fraction diagram is executed in a clockwise direction if and only if $\phi$ is positive; i.e., the disk radius increases after the instantaneous accretion rate (and disk temperature) increases.  Conversely, the loop is executed in an anticlockwise direction if and only if $\phi$ is negative, and the increase in temperature during an outburst lags the increase in disk radius.  As we show in Figures \ref{fig:outburst_ed_ratio6} \& \ref{fig:outburst_ed_ratio3}, the hysteretic loop in the eclipse fraction diagram is executed anticlockwise in both the outburst and superoutburst of Z Cha, leading us to conclude that the disk begins to increase in size before it increases in temperature in both cases.
\par This behaviour can also be seen by considering points \textbf{A} and \textbf{B} in Figure \ref{fig:toymodel}.  By definition, at both of these points, $r_{\rm disk}=r_{\rm ecl}$.  As the out-of-eclipse flux at \textbf{B} is greater than at \textbf{A}, it follows that $\dot{m}$ must also be greater at \textbf{B} than at \textbf{A}.  Assuming the functional forms $\dot{m}(t)$ and $r_{\rm disk}(t)$ are similar, this also implies that an increase in accretion rate lags after an increase in disk size.  This in turn suggests that, at the onset of an outburst, the increase in viscosity in the accretion disk first causes the disk to expand.  The matter in this expanded region of the outer disk then accretes inwards, increasing the mean mass transfer rate within the disk and raising the luminosity of the disk surface.

\subsection{Superhump Period \& Evolution}
\par Our results have implications for the behaviour of a dwarf nova accretion disk during outbursts and superoutbursts, particularly during the onset and the decay of these features.  As the brightening events associated with the superhump are large in this system, it is possible to use the path that they traces in a flux-phase plot (e.g. Figure \ref{fig:flux-phase2}) to estimate how the superhump frequency changes over time.  The gradient of a straight line $m$ in a flux-phase plot with a $y$-axis in units of cycles corresponds to an event recurring with a period $P_{\rm SH}$:
\begin{equation}
P_{\rm SH}=P_{\rm fold}\left(\frac{1}{m}+1\right)=P_{\rm fold}\left(\bar{m}+1\right)
\label{eq:1}
\end{equation}
where $P_{\rm fold}$ is the folding period used to obtain the flux-phase plot and $\bar{m}$ is defined as $1/m$, and is equal to the rate of change of phase as a function of cycle number.  Consequently, the rate of change of $P_{\rm SH}$ with respect to time can be calculated as:
\begin{equation}
\dot{P}_{\rm SH}=\frac{1}{P_{\rm fold}}\frac{\mathrm{d}P_{\rm SH}}{\mathrm{d}N}=\frac{\mathrm{d}\bar{m}}{\mathrm{d}N}
\end{equation}
where $N$ is the time in units of number of cycles since the approximate onset of the superhump at BJD 2458385.  By fitting a parabola to the feature caused by the superhump in a flux-phase diagram of the Sector 3 lightcurve, we find a value $\frac{\mathrm{d}\bar{m}}{\mathrm{d}N}=-8.8(1)\times10^{-5}$, corresponding to $\dot{P}_{\rm SH}=-7.6(1)$\,s/day; the best-fit values for this parabola are given in Table \ref{tab:parab}, and we plot it in Figure \ref{fig:flux-phase2}.
\par At the approximate onset of the superhump, $N=0$ and, using the values in Table \ref{tab:parab} and Equation \ref{eq:1}, the instantaneous superhump period can be calculated to be 0.07774(6)\,d at this time.

\begin{table}
\centering
\begin{tabular}{lll}
\hline
\multicolumn{3}{c}{$\phi_{\rm SH}(N)=aN^2+bN+c$}\\
\hline
\hline
&Value&Error\\
\hline
$a$&$-4.4\times10^{-5}$&$5.0\times10^{-7}$\\
$b$&$5.5\times10^{-3}$&$7.4\times10^{-5}$\\
$c$&$3.4\times10^{-1}$&$2.4\times10^{-3}$\\
\hline
\hline
\end{tabular}
\caption{A table of best-fit parameters, in units of phase/cycle, for a parabola fit to the curve traced by the superhump in Figure \ref{fig:flux-phase2}.  $\phi_{\rm SH}$ is the expected phase of the peak of the superhump, and $N$ is the number of cycles since BJD 2458385, the approximate time of onset for the superhumps.  $P_{\rm fold}=0.0771892$.}
\label{tab:parab}
\end{table}

\par
\begin{figure}
    \includegraphics[width=\columnwidth, trim = 2mm 10mm 12mm 5mm]{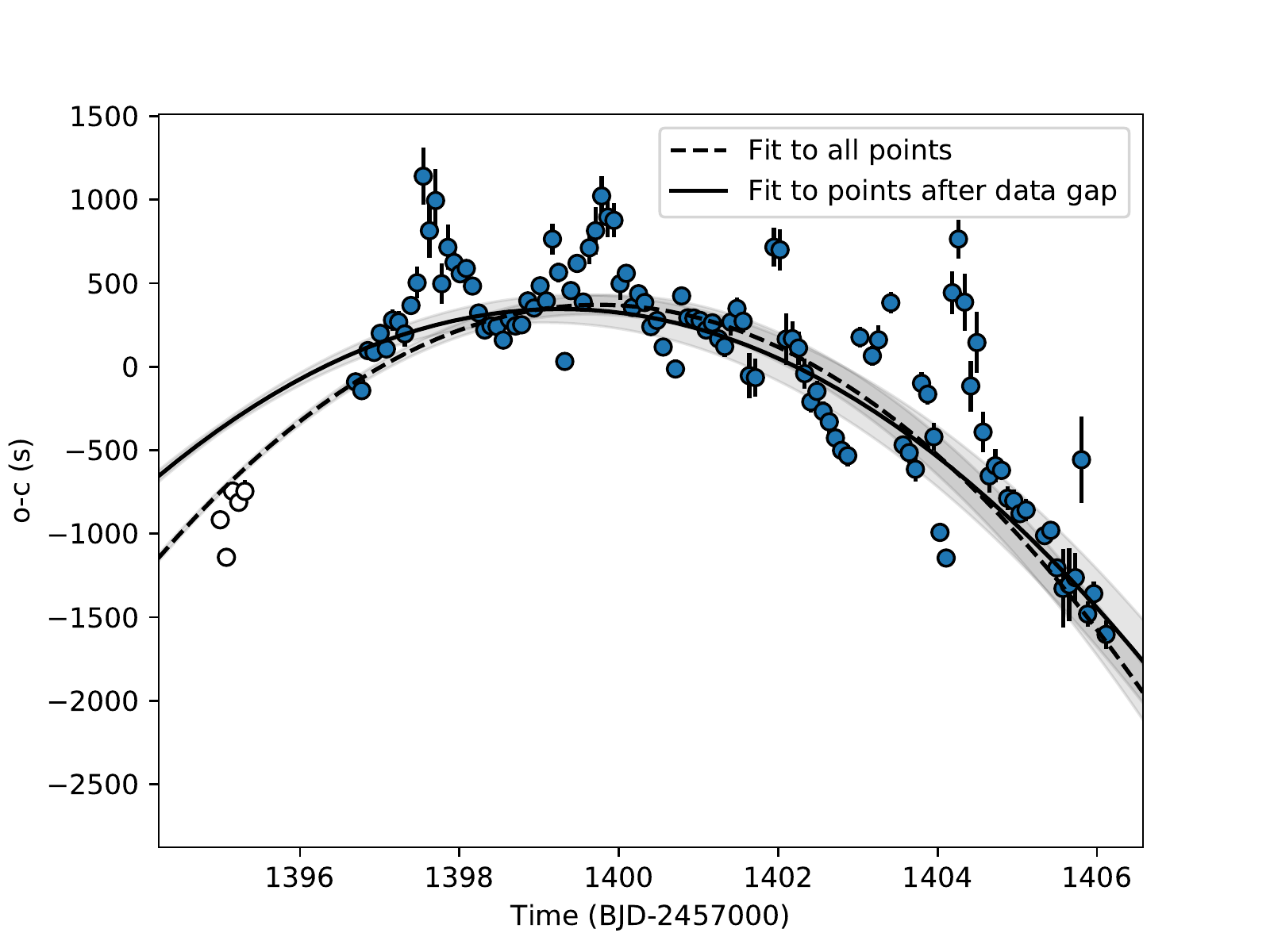}
    \captionsetup{singlelinecheck=off}
    \caption{An O-C diagram of superhump arrival times against a linear ephemeris with a period of 0.0771892\,d and arbitrary phase.  The dashed line is a quadratic ephemeris fit to all superhumps in our sample, corresponding to the quadratic fit to the flux-phase diagram presented in Table \ref{tab:parab}.  The solid line is a quadratic ephemeris fit only to those superhumps which occurred after the large data gap in Sector 3 (in blue).  In each case, 1\,$\sigma$ confidence intervals are highlighted in grey.  Note that this ephemeris significantly overestimates the O-C values of superhumps which occurred before the data gap (in white).}
   \label{fig:sh_oc}
\end{figure}

 The drift in the superhump period can also be investigated using O-C diagrams.  To estimate the arrival time of each superhump, we divide the eclipse-removed lightcurve into segments of 0.0771892\,d, a value close to the approximate median superhump reccurence time.  A Gaussian was then fit to each of these segments to extract the time at which the peak of each superhump occurred.  We compared these arrival times against those predicted by a linear ephemeris with a period of 0.0771892\,d; we show the resultant O-C diagram in Figure \ref{fig:sh_oc}, showing the quadratic ephemeris corresponding to the parabolic fit to the flux-phase diagram presented in Table \ref{tab:parab}.  To check for consistency, we also fit a quadratic ephemeris to only the superhumps which occurred after the large data gap in Sector 3 (shown in blue in Figure \ref{fig:sh_oc}).  When extrapolated, this new ephemeris significantly overestimates the O-C values of superhumps which occurred before the data gap (shown in white in Figure \ref{fig:sh_oc}).  This implies that the mean rate of period decay during and before the data gap was larger than the mean rate of period decay after the data gap.  It is therefore unlikely that the superhump period was constant for any significant length of time during this data gap.

\par A decreasing superhump period has been previously noted in a number of other systems \citep[e.g.][]{Uemura_SPdot}.  \citealp{Kato_Shumps1} found that the evolution of the superhump period in most AWD superoutbursts can be described in three evolutionary `stages':

\begin{itemize}
\item Stage A, in which the superhump period is high and stable.  The superhump period in this stage higher than in both the other stages.
\item Stage B, which occurs after Stage A, in which the superhump period usually decreases smoothly over time.  In some sources, the superhump period may increase instead during this stage.
\item Stage C, which occurs after Stage B, in which the superhump period is low and stable.
\end{itemize}

Our results suggest that the superhump period of Z Cha continued to decrease during the data gap in Sector 3.  As such, we find that the superhump in the 2018 superoutburst of Z Cha was in evolutionary Stage B for the duration of the observation with \textit{TESS}, and neither Stages A nor C were observed.  As the start of the outburst was observed, we are able to state that Stage A did not occur during this superoutburst, but it is unclear whether Stage C occurred as the late stages of the outburst were not observed.  Previous studies of the superhump period in superoutbursts of Z Cha \citep[e.g.][]{Kato_Shumps7} have been unable to determine the period derivative during Stage B using O-C fitting techniques.  As such our measurement calculated using the flux-phase diagram is, as far as we are aware, the first superhump period derivative calculated for Z Cha.
\par Stage A superhumps are interpreted as being the dynamic procession rate at the radius of the 3:1 resonance with the binary orbit.  Stage B superhumps appear later, when a pressure-driven instability in the disk becomes appreciable.  The absence of Stage A superhumps has been noted in a few other SU UMa type AWDs (e.g. QZ Vir and IRXS J0532, \citealp{Kato_Shumps1}), but the physical parameters which determine whether Stage A superhumps will occur in a given superoutburst remain unclear.  Notably, a previous superoutburst of Z Cha in 2014 did show Stage A superhumps \citep{Kato_Shumps7}, indicating that the absence of Type A superhumps is a property related to individual superoutbursts rather than to the system.
\par Additionally, using our estimate of the superhump period at its onset, it is possible to obtain an independent estimate for the mass ratio $q$ of the components of the Z Cha system, where $q$ is the mass ratio defined as:
\begin{equation}
q \equiv M_{\rm d}/M_{\rm a}
\end{equation}
Where $M_{\rm a}$ is the mass of the accretor (in this case the white dwarf) and $M_{\rm d}$ is the mass of the donor star.  We use the empirical relation between $q$ and $\epsilon$ \citep{Knigge_Relation}:
\begin{equation}
q=(0.114\pm0.005)+(3.97\pm0.41)\times(\epsilon-0.025)
\label{eq:exrel}
\end{equation}
where $\epsilon$, the superhump period excess, is a function of superhump period $P_{\rm sh}$ and orbital period $P_{\rm orb}$:
\begin{equation}
\epsilon\equiv\frac{P_{\rm sh}-P_{\rm orb}}{P_{\rm orb}}
\end{equation}

From Equation \ref{eq:exrel}, we thus obtain $q=0.187\pm0.013$.  This mass ratio is consistent with recent estimates of $q$ obtained via independent methods (e.g. \citealp{McAllister_Zeclipses}).

\section{Conclusions}

\par We have performed a study of the timing properties of Z Cha during the \textit{TESS} observations of that source in 2018, as well as a study of how the eclipse properties varied throughout that time period.  We calculate the arrival times of eclipses in this system, and use these results to confidently rule out a number of ephemerides constructed by previous studies.  We thus create a new orbital ephemeris for the Z Cha system, implying the existence of a third body (consistent with previous studies of this object) and finding a new orbital period for this body of $37.5\pm0.5$\,yr.
\par We also study the properties of the `positive superhump'; an oscillation with a period slightly greater than the orbital period which has been observed during superoutbursts in a number of dwarf nova AWDs.  We find that the period associated with the superhump changes significantly during the superoutburst, decreasing towards the orbital period of the system as the superoutburst progresses, and find the rate at which this period decays for the first time in Z Cha.  Notably, we find that the superhump in Z Cha evolves in a non-standard way, skipping evolutionary Stage A entirely.  Superhumps during previous outbursts of this source have been observed to evolve normally, suggesting that the absence of Stage A evolution is a property of an individual superoutburst rather than of an AWD system.
\par Finally, we trace how the depth of an eclipse, and the out-of-eclipse flux, vary over the course of both an outburst and a superoutburst.  We find that, during quiescence, these parameters follow a 1:1 relationship.  Out of quiescence, eclipses deviate from this relationship such that eclipse depth becomes significantly less than out-of eclipse flux.  We interpret this as being due to the quiescent disk being comparable in radius to the eclipsing red dwarf, and hence able to be totally or near-totally eclipsed, whereas the disk during outburst becomes larger and is only partially eclipsed.  We also find evidence of hysteresis in this parameter space, and show that this can be explained by allowing a lag to exist between an increase in the radius of the accretion disk and the instantaneous mass transfer rate within the disk.  We show that this lag is positive, suggesting that the disk first grows in size during an outburst, and then this triggers an increase in the mass transfer rate which causes a heating of the disk.

\section*{Acknowledgements}


\par In this study, we make use of the Numpy, Scipy \citep{Numpy} and Astropy \citep{Astropy} for Python.  Figures in this paper were produced using MatplotLib \citep{Hunter_MatPlotLib} and Inkscape (\url{https://inkscape.org}).
\par C.K. and N.C.S. acknowledge support by the Science and Technology Facilities Council (STFC), and from STFC grant ST/M001326/1. M.K. is funded by a Newton International Fellowship from the Royal Society. P.S. acknowledges support by the National Science Foundation (NSF) grant AST-1514737.




\bibliographystyle{mnras}
\bibliography{/home/jamie/Documents/Bibliographies/refs}








\bsp	
\label{lastpage}
\end{document}